\documentclass[12pt]{article}

\usepackage{epsfig,latexsym,amsfonts,amsmath,amsthm,amssymb,amsbsy,multirow,slashed,wasysym,textcomp,dsfont,comment,mathtools,cancel,cite,datetime,appendix,BOONDOX-calo}
\usepackage{tocloft}
\usepackage{hyperref}

 \newcommand{\badat}{\begin{alignedat}}
 \newcommand{\eadat}{\end{alignedat}}
 \newcommand\scalemath[2]{\scalebox{#1}{\mbox{\ensuremath{\displaystyle #2}}}}
 \def\be{\begin{equation}}
\def\ee{\end{equation}}

\def\p{\partial}

\usepackage{color}

\newcommand{\pink}[1]{\textcolor{\pink}{#1}}

\definecolor{dblue}{rgb}{0.2,0.50,0.80}

\def\Q{\mathcal{Q}}

\def\bh{{\bar h}}
\def\bz{{\bar z}}
\def\bw{{\bar w}}

\def\ba{{\bar a}}
\def\bb{{\bar b}}
\def\bc{{\bar c}}
\def\bd{{\bar d}}
\def\bm{{\bar m}}

\def\tA{\widetilde A}
\def\tF{\widetilde F}
\def\th{\widetilde h}

\def\tC{\widetilde C}

\def\th{\widetilde h}

\def\tvarphi{\widetilde{\varphi}}

\def\tF{\widetilde{F}}
\def\tg{\widetilde{g}}

\def\tR{\widetilde{R}}

\def\KSvec{\mathcal{m}}
\def\tKSvec{\widetilde{\mathcal{m}}}

\numberwithin{equation}{section} 

\begin{document}

 \begin{titlepage}
  \thispagestyle{empty}
  \begin{flushright}
  CPHT-RR095.122020
  \end{flushright}
  \bigskip
  \begin{center}

      \baselineskip=13pt {\LARGE \scshape{Shifting Spin on the Celestial Sphere }}
  
      \vskip1cm 

   \centerline{ 
   {Sabrina Pasterski}${}^\diamondsuit{}$
   {and Andrea Puhm}${}^\blacklozenge{}$
   }

\bigskip\bigskip
 
 \centerline{\em${}^\diamondsuit$ Princeton Center for Theoretical Science, Princeton, NJ 08544, USA}
 
\bigskip
 
\centerline{\em${}^\blacklozenge$   CPHT, CNRS, Ecole Polytechnique, IP Paris, F-91128 Palaiseau, France}

\smallskip

\bigskip\bigskip

\end{center}

\begin{abstract}
  \noindent We explore conformal primary wavefunctions for all half integer spins up to the graviton.  Half steps are related by supersymmetry, integer steps by the classical double copy. 
  The main results are as follows:  we 1) introduce a convenient spin frame and null tetrad to organize all radiative modes of varying spin; 2) identify the massless spin-$\frac{3}{2}$ conformal primary wavefunction as well as the conformally soft Goldstone mode corresponding to large supersymmetry transformations; 3) indicate how to express a conformal primary of arbitrary spin in terms of differential operators acting on a scalar primary; 4) demonstrate that conformal primary metrics satisfy the double copy in a variety of forms -- operator, Weyl, and Kerr-Schild -- and are exact, albeit complex, solutions to the fully non-linear Einstein equations of Petrov type {\bf N}; 5) propose a novel generalization of conformal primary wavefunctions; and 6) show that this generalization includes a large class of physically interesting metrics corresponding to ultra-boosted black holes, shockwaves and more.
 
\end{abstract}

\end{titlepage}

\tableofcontents

\section{Introduction}

Scattering in four-dimensional asymptotically flat spacetimes obeys an infinite-dimensional symmetry algebra that matches the structure of a two-dimensional conformal field theory (CFT) living on the celestial sphere.  In practice, mapping $\cal{S}$-matrix elements to celestial CFT correlators is an integral transform of on-shell momenta. In principle, it presents a new paradigm for identifying and organizing universal features of scattering.  

The goal of the Celestial Holography program is to push beyond kinematics and gain insight into quantum gravity in the bulk asymptotically flat spacetime from the celestial boundary theory.  Over the past few years, important strides have been made towards building a holographic dictionary.  Beyond matching soft memory modes to currents~\cite{Strominger:2017zoo}, and boost eigenstates to local operators~\cite{Pasterski:2016qvg,Pasterski:2017kqt,Pasterski:2017ylz}, recent advances are beginning to translate known universal features of amplitudes into this new representation.
Soft theorems map to factorization theorems at special conformal dimensions~\cite{Nandan:2019jas,Pate:2019mfs,Adamo:2019ipt,Puhm:2019zbl,Guevara:2019ypd,Fan:2019emx}, collinear limits are captured by operator product expansions~\cite{Fan:2019emx,Fotopoulos:2019tpe,Pate:2019lpp,Fotopoulos:2019vac,Banerjee:2020kaa,Fan:2020xjj}, double copy relations in amplitudes persist~\cite{Casali:2020vuy}, and the UV/IR mixing intrinsic to this map has offered more insight into analyticity constraints and a new perspective on renormalization~\cite{Arkani-Hamed:2020gyp}.

The goal of this paper is to expand the existing framework surrounding conformal primary states~\cite{Pasterski:2017ylz,Donnay:2018neh,Donnay:2020guq} and connect it to current progress being made in adjacent subfields so that we will, ultimately, be prepared to apply their tools to our problems.  We take inspiration from recent successes connecting classical observables to on-shell quantum scattering amplitudes~\cite{Cristofoli:2020hnk,Arkani-Hamed:2019ymq} as well as the so-called double copy relation between gravity and gauge theory amplitudes~\cite{Bern:2010ue,Bern:2010yg,Feng:2010my} and its classical counterpart~\cite{Monteiro:2014cda,Luna:2018dpt,Huang:2019cja,Alawadhi:2019urr,White:2020sfn}. In particular, we will exploit the machinery of the classical double copy - of Weyl and Kerr-Schild type - between solutions in gauge theory and gravity.

Our results are organized around the theme of `shifting spin,' which takes on a double meaning.  On the one hand, we will be examining how to step between radiative modes of different spin.  On the other, we will be considering wavefunctions with conformal spins that differ from the bulk helicity of the corresponding field. We demonstrate that within the existing formalism for conformal primary wavefunctions we can construct exact solutions which may serve as backgrounds on top of which to perform perturbative scattering.  We then extend this formalism to encompass a larger class of interesting bulk states. The objective is to set up a framework that will help us explore bulk physics in a manner that may be overlooked if we rely solely on a Mellin transform of perturbative amplitudes.

This paper is organized as follows.  In section~\ref{pre}, we set up a convenient null tetrad and spin frame to describe conformal primary wavefunctions.  In section~\ref{pwtocpw}, we use this to reorganize the presentation of conformal primaries.  Starting from a review of each Mellin representative of spin $s\in\frac{1}{2}\mathbb{Z}$ for $0\le s\le 2$ in section~\ref{Mellinrep}, we then use our spin frame to promote these to conformal primaries in section~\ref{supercpw}.  This allows us to quickly write down a series of interesting new results: identifying the full spin-$\frac{3}{2}$ conformal primary wavefunction, the conformal Goldstone mode for its large gauge symmetry corresponding to local supersymmetry, the generic all-spins expansion of conformal primary radiative modes, and charge operators for these modes (pre-renormalization; counterterms are examined in~\cite{Donnay:2020guq} and~\cite{PPP}).  

Section~\ref{Mellinspin} explains how one can jump between different spins using differential operators.  It begins, in section~\ref{susyq}, with a review of the supersymmetry action on Mellin-transformed amplitudes introduced by~\cite{Fotopoulos:2020bqj}, which we generalize in a manner that allows us to shift the spin of a conformal primary wavefunction by arbitrary half-integer steps.  We then show how one can also jump by integer spin with an operator double copy for curvatures in section~\ref{sec:CDCops}.  One of the focuses of this paper is to consider exact vs perturbative backgrounds.  While the gauge equivalence of perturbative Mellin modes and conformal primary modes would allow the use of amplitudes constructed from either, section~\ref{gaugequiv} shows how this gauge equivalence persists for the finite perturbations.  This will make it easier to extract scattering on finite conformal primary backgrounds from amplitudes methods.

In section~\ref{sec:CDCWeyl}, we use our spin frame to demonstrate that spin-0, spin-1, and spin-2 conformal primary wavefunctions obey the Weyl double copy relations~\cite{Luna:2018dpt}.  This makes manifest the \text{(anti-)} self duality of these solutions.  In section~\ref{sec:CDCKS}, we show that the conformal primary solutions satisfy the Kerr-Schild double copy.  This implies that spin-2 conformal primary wavefunctions are fully non-linear solutions to the Einstein equations.  We examine the Kerr-Schild doubly copy for conformal primary wavefunctions, for their shadow transforms,  and for conformally soft modes in sections~\ref{sec:nonsh}-\ref{sec:cscpw}, respectively, both verifying that these modes become pure gauge when expected and identifying their Petrov type.

At this point, we have a set of nontrivial background configurations with definite conformal weight and spin on which to consider perturbative scattering.  We find that we can expand this set to even more useful backgrounds if we relax our definition of conformal primary wavefunctions.  In section~\ref{sec:gencpw}, we generalize the construction of~\cite{Pasterski:2017kqt} to include non-radiative wavefunctions of definite conformal weight.  We perform a classification for each integer spin in sections \ref{sec:gencpws}-\ref{sec:gencpwt}.  Finally, we apply this classification to a series of interesting backgrounds in section~\ref{sec:bhstates}.  These include boosted black holes, shock wave configurations, and other vacuum-to-vacuum transitions that would otherwise be excluded from a conformal primary analysis.  Celebrated as exact solutions when they were discovered in the 70's and 80's~\cite{Aichelburg:1970dh,Ferrari:1988cc,Dray:1985ie}, these metrics have resurfaced recently in the amplitudes literature.  Identifying these metrics as generalized conformal primaries opens up new opportunities to apply these same amplitudes methods to explore non-perturbative bulk physics in the celestial CFT.

\section{Null Tetrad and Spin Frame}\label{pre}

In this section, we set up a spin frame and null tetrad for the Minkowski metric that will be convenient for our discussion of conformal primary wavefunctions. We use the spinor conventions of~\cite{Elvang:2013cua} since they are also in the mostly-plus signature convention.
In the Celestial Conformal Field Theory  dictionary~\cite{Pasterski:2016qvg,Pasterski:2017kqt,Pasterski:2017ylz}, 2D CFT operators at a point $(w,\bw)$ correspond to 4D bulk wavefunctions defined in terms of a reference direction
\be
q^\mu=(1+w\bw,w+\bw,i(\bw-w),1-w\bw)\,.
\ee
Under an $SL(2,\mathbb{C})$ M\"{o}bius transformation of the celestial sphere
 \be\label{mobius}
 w\mapsto \frac{a w+b}{cw+d}\,,~~~\bw\mapsto \frac{{\bar a} \bw+{\bar b}}{{\bar c}\bw+{\bar d}}\,,
 \ee
with $ad-bc=\ba\bd-\bb\bc=1$, this reference direction transforms as
 \be\label{qtra}
 q^\mu\mapsto |cw+d|^{-2}\Lambda^{\mu}_{~\nu} q^\nu \,,
 \ee
where $\Lambda^{\mu}_{~\nu}$ is the corresponding vector representation of $SO(1,3)\cong SL(2,\mathbb{C})$ (see e.g.~\cite{Oblak:2015qia,Narayanan:2020amh})
\begin{equation}
    \Lambda^\mu_{~\nu}=\frac{1}{2}
    \scalemath{0.83}{
    \left(\begin{array}{cccc}
    a\ba+b\bb+c\bc+d\bd & a\bb+\ba b+\bc d+c\bd & i(a\bb-\ba b+c \bd-\bc d) & -a\ba+b\bb-c\bc+d\bd\\
    a\bc+\ba c+b\bd+\bb d & a \bd+\ba d +b\bc+\bb c & i(a\bd -\ba d -b\bc+\bb c) & -a\bc -\ba c+b\bd+\bb d\\
    i(-a\bc +\ba c -b \bd +d\bb ) & i(-a\bd +\ba d-b \bc+\bb c) & a\bd+\ba d-b\bc -\bb c & i(a\bc -\ba c-b \bd+\bb d\\
    -a \ba- b\bb + c\bc +d\bd & -a \bb- \ba b+c\bd+\bc d & i(-a\bb+\ba b+c\bd -\bc d) & a\ba -b\bb -c\bc +d\bd
    \end{array}\right)\,.
    }
\end{equation}
From this reference direction, one can naturally construct two polarization vectors
\be\label{eq:pol}
\epsilon_+^\mu=\frac{1}{\sqrt{2}}\p_w q^\mu,~~~\epsilon_-^\mu=\frac{1}{\sqrt{2}}\p_{\bw} q^\mu\,,
\ee
which obey the following inner products 
\be\label{qepsIDs}
q\cdot \epsilon_J=0\,, \quad q^2=\epsilon_J^2=0\,, \quad\epsilon_J\cdot \epsilon_{-J}=1\,,
\ee
for $J=\pm1$.  

We see that we are one null vector shy of a tetrad for Minkowski space.  However, we have one more four vector at our disposal for a spacetime wavefunction: $X^\mu.$ Let us now proceed to construct a tetrad in terms of $\{X^\mu,q^\mu,\epsilon_+^\mu,\epsilon_-^\mu\}$, which form a basis for the tangent space for generic $X^\mu$.  We would also like to demand that vectors in our tetrad transform covariantly under $SL(2,\mathbb{C})$.  A tetrad that fits these criteria is given by
\be\label{tetrad}
l^\mu=\frac{q^\mu}{-q\cdot X}\,, ~~~n^\mu=X^\mu+\frac{X^2}{2}l^\mu\,, ~~~m^\mu=\epsilon^\mu_++(\epsilon_+\cdot X) l^\mu\,, ~~~\bar{m}^\mu=\epsilon^\mu_-+(\epsilon_-\cdot X) l^\mu\,.
\ee
One can check that our tetrad obeys the standard normalization conditions
\be\label{eq:iptetrad}
l\cdot n=-1\,,~~m\cdot\bar{m}=1\,,~~~l^2=n^2=m^2={\bar m}^2=0\,,~~l\cdot m=n\cdot m=0\,,~~{\bar{m}}^\mu=(m^\mu)^*\,,
\ee 
and transforms covariantly under $SL(2,\mathbb{C})$ as follows:\footnote{The combination of $\epsilon^\mu_+$ and $q^\mu$ appearing is such that $m^\mu$ not only transforms covariantly under Lorentz transformations, but also is invariant under shifts $\epsilon_+ \mapsto\epsilon_+ +\alpha q$, with the same being true for $\epsilon_+\leftrightarrow\epsilon_-$ and $m\leftrightarrow\bar{m}$.  Allowing spacetime dependence and demanding conformal covariance gives rise to a natural spacetime-dependent polarization vector independent of the standard momentum space gauge ambiguity. Furthermore, the normalization condition~\eqref{eq:iptetrad} prevents us from shifting the weights of $m$ and ${\bar m}$. \label{gaugechange}}
\be\label{tetradcov}
l^\mu\mapsto \Lambda^{\mu}_{~\nu} l^\nu\,,~~n^\mu\mapsto \Lambda^{\mu}_{~\nu} n^\nu\,,
~~m^\mu\mapsto \frac{cw+d}{{\bar c}\bw+{\bar d}}\Lambda^{\mu}_{~\nu} m^\nu\,,~~{\bar m}^\mu\mapsto \frac{{\bar c}\bw+{\bar d}}{cw+d}\Lambda^{\mu}_{~\nu} {\bar m}^\nu\,.
\ee
An operator or wavefunction is said to have $SL(2,\mathbb{C})$ conformal dimension $\Delta$ and spin $J$ if
\be\label{deltaj}
\mathcal{O}_{\Delta,J}\Big(\Lambda^{\mu}_{~\nu} X^\nu;\frac{a w+b}{cw+d},\frac{{\bar a} \bw+{\bar b}}{{\bar c}\bw+{\bar d}}\Big)=(cw+d)^{\Delta+J}({\bar c}\bw+{\bar d})^{\Delta-J}D(\Lambda)\mathcal{O}_{\Delta,J}(X^\mu;w,\bw)\,,
\ee
where $D(\Lambda)$ is the representation of the corresponding Lorentz transformation appropriate for the $3+1$D indices of the operator $\mathcal{O}$. We thus see that $l^\mu$ and $n^\mu$ are real four vectors with $SL(2,\mathbb{C})$  conformal weights $(h,\bh)\equiv\frac{1}{2}(\Delta+J,\Delta-J)=(0,0)$, whereas $m^\mu$ and ${\bar m}^\mu$ are complex four vectors with conformal dimension $\Delta=0$ and spin $J=\pm 1$, respectively.

The flat metric can be written in terms of our tetrad~\eqref{tetrad} as 
\be
\eta^{\mu\nu}=-l^\mu n^\nu-n^\mu l^\nu+m^\mu {\bar m}^\nu+{\bar m}^\mu m^\nu\,.
\ee
We can further decompose the elements of this tetrad into a spin frame.  We want
\be\label{spinframe}
l_{a{\dot b}}=o_a\bar{o}_{\dot b}\,,~~n_{a{\dot b}}=\iota_a\bar{\iota}_{\dot b}\,,~~m_{a{\dot b}}=o_a{\bar\iota}_{\dot b}\,,~~\bar{m}_{a{\dot b}}=\iota_a{\bar o}_{\dot b}\,,
\ee
where for a four vector $v^\mu$ we go between spinor and vector indices via~\cite{Elvang:2013cua}
\be
v_{a{\dot b}}=v_\mu(\sigma^\mu)_{a\dot b}\,,~~~v^\mu=-\frac{1}{2}\mathrm{tr}(v\bar{\sigma}^\mu)\,,
\ee
where
\be\label{ivdw}
(\sigma^\mu)_{a\dot b}=(\mathds{1},\sigma^i)_{a\dot b}\,,~~(\bar{\sigma}^\mu)^{{\dot a} b}=(\mathds{1},-\sigma^i)^{{\dot a} b}\,.
\ee
Here, the undotted (dotted) indices refer to left (right) handed $SL(2,\mathbb{C})$ spinors and these indices are raised and lowered with
\be\label{lc2}
\varepsilon^{ab}=\varepsilon^{\dot{a}\dot{b}}=-\varepsilon_{ab}=-\varepsilon_{\dot{a}\dot{b}}=\left(\begin{array}{cc}
0&1\\-1&0\\
\end{array}\right).
\ee
We then see that the decomposition~\eqref{spinframe} holds for
\be\label{spinframespinors}
o_a=\sqrt{\frac{2}{q\cdot X}}\left(\begin{array}{c} \bw \\-1 \end{array}\right)\,,~~~\iota_a=\sqrt{\frac{1}{q\cdot X}}\left(\begin{array}{c}t-z-w(x-iy) \\-x-iy+w(t+z) \end{array}\right)
\ee
up to an overall phase ambiguity which we fix by setting ${\bar o}_{\dot a}=(o_a)^*$ and ${\bar \iota}_{\dot a}=(\iota_a)^*$ in the region where $q\cdot X>0$ and analytically continue from there.
The spinor $o_a$ has a simple relation to the spinor helicity variables 
\be\label{oqspinorhelicity}
|q]_a=\sqrt{ q\cdot X}o_a\,,~~~\langle q|_{\dot a}=\sqrt{ q\cdot X}{\bar o}_{\dot a}\,,~~~q_{a\dot{a}}=-|q]_a\langle q|_{\dot a}\,.
\ee
We also have
\be\label{lcr}
o_a \iota_b-\iota_a o_b=\sqrt{2}\varepsilon_{ab}\,,
\ee
as well as the simple linear relation\footnote{We are essentially using conformal covariance to pick a reference spinor. 
From an amplitudes perspective, this would correspond to using a differential operator in place of an arbitrary reference spinor, replacing $X^\mu$ with $i\p_{k^\mu}$ for each external leg, or an appropriate Mellin basis analog. 
The linear relationship between $\iota$ and $\bar{o}$ is reminiscent of the incidence relation defining the embedding of Minkowski space into twistor space $Z^\alpha=(\omega^A,\pi_{A'})$ as the locus~\eqref{inc}, $\omega^A=iX^{AA'}\pi_{A'}$ so long as the contour integral in the Penrose transform localizes to $\pi_{A'}\mapsto {\bar o}_{\dot a}$, $\omega^A \mapsto i\sqrt{2}\iota^a$.
It would be interesting to connect our classical celestial double copy, in particular section~\ref{sec:CDCWeyl}, to the recent results of~\cite{White:2020sfn} showing that the Weyl double copy can be derived from twistor theory.  We leave fleshing out these details to future work.  } 
\be\label{inc}
\iota_a=\frac{1}{\sqrt{2}}X_{a\dot{b}} \bar{o}^{\dot b}\,,
\ee
where 
\be
X_{a\dot{b}}=\left(\begin{array}{cc}
-t+z & x-iy\\ x+iy & -t-z
\end{array}\right)
\ee
is the matrix representation of the spacetime position vector.

We close this preliminary section by examining the $SL(2,\mathbb{C})$ covariance of our spinors $\{o_a,\iota_a\}$.  Because of~\eqref{inc} and the covariance properties of the Infeld-van der Waerden symbols~\eqref{ivdw} and the antisymmetric tensor~\eqref{lc2}, we need only perform an explicit transformation of $o_a$. Under a M\"{o}bius transformation
\be
o_a\mapsto (cw+d)^{\frac{1}{2}}({\bar c}\bw+{\bar d})^{-\frac{1}{2}}(Mo)_a\,,
\ee
where 
\be\label{eq:Mabcd}
M=\left(\begin{array}{cc} \ba& -\bb \\ -\bc &\bd \end{array}\right)
\ee
is an element of $\overline{ SL(2,\mathbb{C})}$.
Meanwhile, from the intertwining relation of the  Infeld-van der Waerden symbols~\eqref{ivdw} we get
\be
X_{a{\dot b}}\mapsto (M X M^\dagger)_{a{\dot b}}=\Lambda^\mu_{~\nu}X^\nu\eta_{\mu\rho}(\sigma^\rho)_{a{\dot b}}\,.
\ee
We thus have
\be
\iota_a\mapsto (cw+d)^{-\frac{1}{2}}({\bar c}\bw+{\bar d})^{\frac{1}{2}}(M\iota)_a\,,
\ee
so that
\be\begin{array}{l|l|r}
& \Delta & J\\
\hline
o_a & 0 & \frac{1}{2} \\
{\bar o}_{\dot a} & 0 & -\frac{1}{2}\\
\iota_a & 0 & -\frac{1}{2} \\
{\bar \iota}_{\dot a} & 0 & \frac{1}{2}\\
\end{array}
\ee
from which we see that the tensor products~\eqref{spinframe} reproduce the correct conformal dimensions for the members of our tetrad.

\section{From Plane Waves to Conformal Primaries}\label{pwtocpw}
This section applies the above framework to reorganize wavefunctions relevant to celestial CFT correlators.  Scattering amplitudes are usually calculated with external states in the plane wave basis, making translation symmetry manifest. The fact that the 4D Lorentz group $SL(2,\mathbb{C})$ acts as the 2D global conformal group on the celestial sphere at null infinity implies the existence of a conformal basis of asymptotic states which makes conformal covariance manifest. For massless\footnote{Massive bosonic wavefunctions were constructed for the scalar in~\cite{Pasterski:2016qvg} and generalized to arbitrary integer spin in~\cite{Law:2020tsg}.  Massive fermionic wavefunctions were constructed in~\cite{Narayanan:2020amh}, and generalized to arbitrary dimensions for spin-$\frac{1}{2}$ in~\cite{Muck:2020wtx}.} particles, this change of basis to so-called {\it celestial amplitudes} is achieved by a Mellin transform in the energy of the external states~\cite{deBoer:2003vf, Cheung:2016iub,Pasterski:2017kqt,Pasterski:2017ylz}.  As shown in~\cite{Fotopoulos:2020bqj}, the Mellin representatives of different spin are related by supersymmetry. It is also straightforward to show that the integer spin examples are related by the classical double copy. We review these massless Mellin wavefunctions which were cataloged for integer spins 0, 1 and 2 in~\cite{Pasterski:2017kqt}, and for half integer spins $\frac{1}{2}$ and  $\frac{3}{2}$ in~\cite{Fotopoulos:2020bqj} in section~\ref{Mellinrep}.

The Mellin representatives for spins \{1, $\frac{3}{2}$, 2\} fail to satisfy~\eqref{deltaj}; but judiciously adding pure gauge terms yields {\it conformal primary wavefunctions} which do transform according to~\eqref{deltaj}. Massless spin-1 and spin-2 conformal primary wavefunctions were introduced in~\cite{Pasterski:2017kqt}.
We believe the explicit form of the massless spin-$\frac{3}{2}$ conformal primary wavefunction we propose in section~\ref{supercpw} is new. Our presentation of conformal primary wavefunctions in terms of the null tetrad and its spin frame reveals that these wavefunctions satisfy double copy relations of the Weyl and Kerr-Schild types which we discuss in sections~\ref{sec:CDCWeyl} and \ref{sec:CDCKS}.

\subsection{Mellin Representatives}\label{Mellinrep}
Recall the linearized equations of motion for massless particles of 4D spin $s=|J|$ in vacuum of relevance to any supergravity theory:
\be\label{lineom}
\begin{array}{lll}
 s=0 &  \mathrm{Klein}$-$\mathrm{Gordon} & \Box \phi=0\\
 s=\frac{1}{2} & \mathrm{Dirac} & \gamma^\mu\p_\mu \psi=0\\
 s=1 & \mathrm{Maxwell} & \Box A_\mu-\p_\mu\p^\nu A_\nu=0\\
 s=\frac{3}{2} & \mathrm{Rarita}$-$\mathrm{Schwinger} & \gamma^{\mu\nu\rho}\p_\nu\chi_\rho=0\\
  s=2 & \mathrm{linearized~Einstein} & \partial_\sigma \partial_\nu h^\sigma_{~\mu}+\partial_\sigma \partial_\mu h^\sigma_{~\nu}-\partial_\mu \partial_\nu h-\Box h_{\mu\nu}=0
\end{array}
\ee
with the linearized metric perturbation given in $g_{\mu\nu}=\eta_{\mu\nu}+h_{\mu\nu}$. Massless particles are labeled by null momenta which, in four dimensions, are described by three parameters: a point on a two-sphere and a scaling given by its energy, $k_\mu=\omega q_\mu(w,\bw)$. The standard procedure for finding scattering solutions to these equations is to start from the scalar solution to the massless Klein-Gordon equation given by the plane wave $e^{\pm ik\cdot X}$ with $k^2=0$ and find the appropriate polarization spinor/tensor to satisfy~\eqref{lineom}. 
The authors of~\cite{Fotopoulos:2020bqj} showed that the standard Mellin map
\be
\mathcal{M}: \mathrm{Plane~Wave~Solution} \mapsto \mathrm{Mellin~Representative}
\ee
where
\be
\mathcal{M}(\cdot)=\int_0^\infty d\omega \omega^{\Delta-1}(\cdot)\,,
\ee
works for for each spin in the supermultiplet, taking the plane wave solutions of~\eqref{lineom} to representatives of the gauge equivalence class of conformal primary wavefunctions of conformal dimension $\Delta$, and spin $J$ equal to the helicity of the pre-image.  Starting from the scalar case
\begin{equation}\label{phiDelta}
\badat{2}
 \phi^{\Delta,\pm}(X;w,\bw)&=\int_0^\infty d\omega \omega^{\Delta-1} e^{\pm i \omega q\cdot X- \varepsilon q^0\omega}=\frac{(\mp i)^\Delta \Gamma(\Delta)}{(-q\cdot X_\pm)^\Delta}\,,
\eadat
\end{equation}
where $X^\mu_\pm=X^\mu\pm i\varepsilon\{-1,0,0,0\}$ is used as a regulator,\footnote{In the spinor definitions from section~\ref{pre}, we take $X\mapsto X_{\pm}$ everywhere to regulate consistently, with the caveat that the spinors $o_a$ and $\iota_a$ are no longer related by complex conjugation to $\bar{o}_{\dot{a}}$ and $\bar{\iota}_{\dot{a}}$, respectively, for finite $\varepsilon$.
}
we can translate the results from~\cite{Fotopoulos:2020bqj} to the spin frame language we introduced above. We arrive at the mapping summarized in table~\ref{tab:MellinReps}.
\begin{table}[t]
\renewcommand*{\arraystretch}{1.3}
    \centering
    \begin{tabular}{l|l|l}
    &Plane Wave Solutions & Mellin Representatives\\
    \hline
    $s=0$ &  $e^{\pm i\omega q\cdot X}$ &$\phi^{\Delta,\pm}$\\
    $s=\frac{1}{2}$ & $\sqrt{\omega}|q]_a e^{\pm i\omega q\cdot X},~~ \pm\sqrt{\omega}\langle q|_{\dot a} e^{\pm i\omega q\cdot X}$ & $o_a^\pm\phi^{\Delta,\pm},~~\pm{\bar o}_{\dot a}^\pm\phi^{\Delta,\pm}$\\
    $s=1$ & $\epsilon^\mu_+ e^{\pm i\omega q\cdot X},~~\epsilon^\mu_- e^{\pm i\omega q\cdot X}$ & $\epsilon^\mu_+\phi^{\Delta,\pm},~~\epsilon^\mu_-\phi^{\Delta,\pm}$\\
    $s=\frac{3}{2}$ & $\sqrt{\omega}|q]_a  \epsilon^\mu_+e^{\pm i\omega q\cdot X},~~\pm\sqrt{\omega}\langle q|_{\dot a} \epsilon^\mu_-e^{\pm i\omega q\cdot X}$  & $o_a^\pm\epsilon^\mu_+\phi^{\Delta,\pm},~~\pm{\bar o}_{\dot a}^\pm\epsilon^\mu_-\phi^{\Delta,\pm}$\\
    $s=2$ & $\epsilon^\mu_+\epsilon^\nu_+ e^{\pm i\omega q\cdot X},~~\epsilon^\mu_-\epsilon^\nu_- e^{\pm i\omega q\cdot X}$   & $\epsilon^\mu_+\epsilon^\nu_+ \phi^{\Delta,\pm},~~\epsilon^\mu_-\epsilon^\nu_- \phi^{\Delta,\pm}$
    \end{tabular}
    \caption{Mellin representatives of spins $s=\{0,\frac{1}{2},1,\frac{3}{2},2\}$.}
    \label{tab:MellinReps}
\end{table}
Note that we have implicitly projected the spinor equations in~\eqref{lineom} onto irreducible Weyl representation solutions, and made use of the properties $|-q]_a=+|q]$ and $\langle -q|_{\dot{a}}=-\langle q|_{\dot{a}}$. We have also reinstated the $\pm i\varepsilon$ regulators for the spinors $o_a$ and $o_{\dot{a}}$.
While the polarization vectors depend solely on $q^\mu(w,\bw)$, the spinors and the scalar wavefunction depend on both $X^\mu$ and $q^\mu(w,\bw)$.

\subsection{Conformal Primaries in the Supermultiplet}\label{supercpw}
From the null tetrad and the spin frame introduced in section~\ref{pre}, it is now straightforward to see how to promote the Mellin transformed plane wave solutions listed in table~\ref{tab:MellinReps} for the various spins to wavefunctions that solve the 4D linearized equations of motion~\eqref{lineom} and transform as 2D conformal primaries~\eqref{deltaj}: substitute $m^\mu$ for any appearance of $\epsilon^\mu_+$, and $\bar{m}^\mu$ for any $\epsilon^\mu_-$. 
For convenience and to match onto the normalization of the conformal primary wavefunctions in~\cite{Pasterski:2017kqt}, we will strip off the prefactor $(\mp i)^\Delta \Gamma(\Delta)$ in~\eqref{phiDelta} and thus define
\be
\varphi^{\Delta,\pm}=\frac{1}{(-q\cdot X_\pm)^\Delta}\,
\ee
as the spin-0 conformal primary wavefunction. Indeed it transforms as a 2D scalar in~\eqref{deltaj}.  In addition, we will add a compensating phase to our definition of the negative helicity fermionic wavefunctions to remove the corresponding $\pm$ signs in table~\ref{tab:MellinReps}.

Spin-1 and spin-2 conformal primary wavefunctions were defined in~\cite{Pasterski:2017kqt} as solutions to the Maxwell and linearized Einstein equations in~\eqref{lineom}, transforming according to~\eqref{deltaj} for $J=\pm1$ and $J=\pm2$, respectively, and were shown to obey the harmonic and radial gauge conditions. The positive helicity wavefunctions are given by
\begin{equation}
    A^{\Delta,\pm}_{\mu;J=+ 1}=m_\mu^\pm \varphi^{\Delta,\pm} \,, \quad h^{\Delta,\pm}_{\mu\nu;J=+2}=m_\mu^\pm m_\nu^\pm \varphi^{\Delta,\pm}\,,
\end{equation}
while the negative helicity ones are 
\begin{equation}
    A^{\Delta,\pm}_{\mu;J=-1}=\bar{m}_\mu^\pm \varphi^{\Delta,\pm} \,, \quad h^{\Delta,\pm}_{\mu\nu;J=-2}=\bar{m}_\mu^\pm \bar{m}_\nu^\pm \varphi^{\Delta,\pm}\,.
\end{equation}

We define spin-$\frac{1}{2}$ and spin-$\frac{3}{2}$ conformal primary wavefunctions as follows, matching the format of~\cite{Pasterski:2017kqt} for the case $d=2$ and treating left and right handed fermionic wavefunctions separately to emphasize the distinction between $3+1$D helicity and $2$D spin.  
For ease of notation, we will suppress spinor indices but leave vector indices explicit, and use bars to denote right handed spinors. 
In the Weyl basis
\be
\gamma^\mu=\left(\begin{array}{cc}0&\sigma^\mu \\ \bar{\sigma}^\mu &0\end{array}\right),
\ee
which we will use to rewrite the spin-$\frac{1}{2}$ and $\frac{3}{2}$ equations from~\eqref{lineom} for each chirality.

~\\
\noindent{\bf Definition:} An outgoing/incoming $(+/-)$, left(right)-handed {\it massless spin-$\frac{1}{2}$ conformal primary Weyl spinor}  $\psi^{\Delta,\pm}_J$ ($\bar{\psi}^{\Delta,\pm}_J$) on $\mathbb{R}^{1,3}$:
\begin{itemize}
\item Is a solution to the Weyl equation
\be \label{12Weyl}
\bar{\sigma}^\mu\p_\mu \psi^{\Delta,\pm}_J=0\,,~~~\sigma^\mu\p_\mu \bar{\psi}^{\Delta,\pm}_J=0\,.
\ee
\item Transforms as a $3+1$-dimensional spinor as well as a 2-dimensional spin-$\frac{1}{2}$ conformal primary of conformal weight $\Delta$ and spin $J$
\begin{equation}
    \badat{2}
\psi^{\Delta,\pm}_J\Big(\Lambda^\mu_{~\nu} X^\nu;\frac{a w+b}{cw+d},\frac{{\bar a} \bw+{\bar b}}{{\bar c}\bw+{\bar d}}\Big)&=(cw+d)^{\Delta+J}({\bar c}\bw+{\bar d})^{\Delta-J}M\psi^{\Delta,\pm}_J(X^\mu;w,\bw)\,,\\
{\bar \psi}^{\Delta,\pm}_J\Big(\Lambda^\mu_{~\nu} X^\nu;\frac{a w+b}{cw+d},\frac{{\bar a} \bw+{\bar b}}{{\bar c}\bw+{\bar d}}\Big)&=(cw+d)^{\Delta+J}({\bar c}\bw+{\bar d})^{\Delta-J}{\bar M} {\bar \psi}^{\Delta,\pm}_J(X^\mu;w,\bw)\,,
\eadat
\end{equation}
where $M$ is the  $(\frac{1}{2},0)$ representation of the Lorentz algebra corresponding to the M\"{o}bius transformation~\eqref{mobius}, given by~\eqref{eq:Mabcd}, and ${\bar M}=(M^{-1})^\dagger$ is the $(0,\frac{1}{2})$ representation.
\end{itemize}  
A complete scattering basis is given by $\Delta=1+i\lambda$.  For these wavefunctions $J=\pm\frac{1}{2}$.
\vspace{2em}

\noindent{\bf Definition:} An outgoing/incoming $(+/-)$, left(right)-handed {\it massless spin-$\frac{3}{2}$ conformal primary Weyl spinor}  $\chi^{\Delta,\pm}_{\mu;J}$ (${\bar\chi}^{\Delta,\pm}_{\mu;J}$) on $\mathbb{R}^{1,3}$:
\begin{itemize}
\item Is a solution to the chiral projection of the Rarita-Schwinger equation
\be\label{32Weyl}
\varepsilon^{\mu\nu\rho\kappa}\bar{\sigma}_\nu\p_\rho \chi^{\Delta,\pm}_{\kappa;J}=0\,,~~~\varepsilon^{\mu\nu\rho\kappa}{\sigma}_\nu\p_\rho {\bar \chi}^{\Delta,\pm}_{\kappa;J}=0\,.
\ee
\item Obeys the harmonic and radial gauge conditions
\be\label{gaugefix}
\p^\mu\chi^{\Delta,\pm}_{\mu;J}=0\,,~~~X^\mu \chi^{\Delta,\pm}_{\mu;J}=0\,,
\ee
and similar expressions for the barred versions. 
\item Transforms as a $3+1$-dimensional spinor as well as a 2-dimensional spin-$\frac{3}{2}$ conformal primary of conformal weight $\Delta$ and spin $J$
\begin{equation}
\badat{2}
\chi^{\Delta,\pm}_{\mu;J}\Big(\Lambda^\mu_{~\nu} X^\nu;\frac{a w+b}{cw+d},\frac{{\bar a} \bw+{\bar b}}{{\bar c}\bw+{\bar d}}\Big)&=(cw+d)^{\Delta+J}({\bar c}\bw+{\bar d})^{\Delta-J}\Lambda_\mu^{~\nu}M\chi^{\Delta,\pm}_{\nu;J}(X^\mu;w,\bw)\,,\\
{\bar\chi}^{\Delta,\pm}_{\mu;J}\Big(\Lambda^\mu_{~\nu} X^\nu;\frac{a w+b}{cw+d},\frac{{\bar a} \bw+{\bar b}}{{\bar c}\bw+{\bar d}}\Big)&=(cw+d)^{\Delta+J}({\bar c}\bw+{\bar d})^{\Delta-J}\Lambda_\mu^{~\nu}{\bar M} {\bar\chi}^{\Delta,\pm}_{\nu;J}(X^\mu;w,\bw)\,,
\eadat
\end{equation}
where $M$ is the  $(\frac{1}{2},0)$ representation of the Lorentz algebra corresponding to the M\"{o}bius transformation~\eqref{mobius}, given by~\eqref{eq:Mabcd}, and $\Lambda$ is the usual vector representation.
\end{itemize}  
A complete scattering basis is given by $\Delta=1+i\lambda$.  For these wavefunctions $J=\pm\frac{3}{2}$.\vspace{1em}

The positive helicity solutions to~\eqref{12Weyl} and~\eqref{32Weyl} are given by the left-handed spinors
\begin{equation}\label{nonshadowLeft}
 \psi^{\Delta,\pm}_{J=+\frac{1}{2}}=o^\pm\varphi^{\Delta,\pm}\,, \quad \chi^{\Delta,\pm}_{\mu;J=+\frac{3}{2}}= o^\pm m_\mu^\pm\varphi^{\Delta,\pm}\,,\\
\end{equation}
while the negative helicity solutions are given by the right-handed spinors
\begin{equation}\label{nonshadowRight}
{\bar\psi}^{\Delta,\pm}_{J=-\frac{1}{2}}= {\bar o}^\pm\varphi^{\Delta,\pm}\,, \quad {\bar\chi}^{\Delta,\pm}_{\mu;J=-\frac{3}{2}}= {\bar o}^\pm {\bar m}_\mu^\pm\varphi^{\Delta,\pm}\,.
\end{equation}
This completes the list of spin $s=\{0,\frac{1}{2},1,\frac{3}{2},2\}$ conformal primary wavefunctions whose 4D helicity $\pm s$ is identified with the 2D spin $J$. Table~\ref{tab:CPWspinors} summarizes the positive helicity conformal primary wavefunctions in spinor notation. The negative helicity wavefunctions are obtained by the replacements $o\mapsto \bar{o}$ and $\bar{\iota}\mapsto \iota$. 
Because the value of $J$ uniquely labels the kind of particle, it is convenient to introduce the compact notation $\Phi_{\Delta,J}(X;w,\bw)$ for 4D solutions to~\eqref{lineom} which transform as 2D spin-$J$ conformal primaries~\eqref{deltaj} with conformal dimension $\Delta=1+i\lambda$. 
Unless necessary we omit the $\pm i \varepsilon$ regulator henceforth.

\begin{table}[ht]
\renewcommand*{\arraystretch}{1.3}
    \centering
    \begin{tabular}{l|l|c}
    $\Phi_{\Delta,J}$&$\mathrm{Wavefunction}$&Reference\\
    \hline
  $\Phi_{\Delta,0}$ & $\varphi^{\Delta}$&\cite{Pasterski:2017kqt}\\
  $\Phi_{\Delta,+\frac{1}{2}}$ & $\psi^{\Delta}_{J=+\frac{1}{2}}=o\varphi^{\Delta}$& \cite{Fotopoulos:2020bqj}\\
  $\Phi_{\Delta,+1}$ & $A_{J=+1}^{\Delta}=o {\bar \iota} \varphi^{\Delta}$& \cite{Pasterski:2017kqt}\\
 $\Phi_{\Delta,+\frac{3}{2}}$ & $\chi^{\Delta}_{J=+\frac{3}{2}}=o o{\bar \iota}\varphi^{\Delta}$&---\\
 $\Phi_{\Delta,+2}$  &  $h_{J=+2}^{\Delta}=o o {\bar \iota} {\bar \iota}\varphi^{\Delta}$&\cite{Pasterski:2017kqt}
 \end{tabular}
 \caption{Positive helicity conformal primary wavefunctions in spinor notation.}
 \label{tab:CPWspinors}
\end{table}

Besides the basis of conformal primary wavefunctions corresponding to $\Phi_{\Delta,J}$, we can construct from their shadow transform a basis of conformal shadow primaries
\begin{equation}\label{ShadowPhi}
\widetilde{\Phi}_{\Delta,J}=\widetilde{{\Phi}_{2-\Delta,-J}}\,,
\end{equation}
whose conformal dimensions are flipped and shifted by 2, while their 2D spins are flipped. We note that while each of the non-shadow modes $\Phi_{\Delta,J}$ has $J=\ell$, where $J$ is the 2D spin and $\ell=\pm s$ is the $3+1$D helicity, the shadow modes $\widetilde{\Phi}_{\Delta,J}$ have $J=-\ell$.
The shadow wavefunctions of spin-0, spin-1 and spin-2 were constructed in~\cite{Pasterski:2017kqt}, and are given by
\begin{equation}
    \widetilde{\varphi}^{\Delta,\pm}=(-X_\pm^2)^{\Delta-1}\varphi^{\Delta,\pm}\,, \quad  \widetilde{A}^{\Delta,\pm}_{\mu;J}=(-X_\pm^2)^{\Delta-1}A^{\Delta,\pm}_{\mu;J}\,, \quad \widetilde{h}^{\Delta,\pm}_{\mu;J}=(-X_\pm^2)^{\Delta-1}h^{\Delta,\pm}_{\mu\nu;J}\,.
\end{equation}
We add to this list the following fermionic wavefunctions, noting that the 4D helicity and the 2D spin are no longer identified. 
The positive helicity conformal shadow primaries are given by the left-handed spinors
\begin{equation}\label{shadowLeft}
\widetilde{\psi}^{\Delta,\pm}_{J=-\frac{1}{2}}=\sqrt{2}\iota^\pm(-X_\pm^2)^{\Delta-\frac{3}{2}}\varphi^{\Delta,\pm}\,, \quad 
\widetilde{\chi}^{\Delta,\pm}_{\mu;J=-\frac{3}{2}}=\sqrt{2} \iota^\pm {\bar m}_\mu^\pm (-X_\pm^2)^{\Delta-\frac{3}{2}}\varphi^{\Delta,\pm}\,,
\end{equation}
while the negative helicity shadow wavefunctions are given by the right handed spinors
\begin{equation}\label{shadowRight}
\widetilde{{\overline\psi}}^{\Delta,\pm}_{J=+\frac{1}{2}}=\sqrt{2}{\bar \iota}^\pm(-X_\pm^2)^{\Delta-\frac{3}{2}}\varphi^{\Delta,\pm}\,, \quad \widetilde{{\overline\chi}}^{\Delta,\pm}_{\mu;J=+\frac{3}{2}}=\sqrt{2} {\bar \iota}^\pm  m_\mu^\pm (-X_\pm^2)^{\Delta-\frac{3}{2}}\varphi^{\Delta,\pm}\,.
\end{equation}
We will demonstrate in a forthcoming paper~\cite{PPP} that the proposed half integer spin solutions~\eqref{nonshadowLeft}-\eqref{nonshadowRight} and~\eqref{shadowLeft}-\eqref{shadowRight} are indeed related by the shadow transformation. 

For now, we postulate that we can generically reach the shadow primaries $\widetilde{\Phi}_{\Delta,J}$ with the same $SL(2,\mathbb{C})$ conformal dimension and spin $(\Delta,J)$ as the primaries $\Phi_{\Delta,J}$ via the map\footnote{We would like to point out the tension between the normalization that grants our spin frame the desired inner products to produce our null tetrad, and the factors of $X^2$ needed for them to satisfy the Dirac equation for the respective Weyl spinors. The spinors are conformal primaries of dimension $0$ and spin $\pm\frac{1}{2}$ in either case, but the spinors which are left handed solutions to the Dirac equation are $\{o_a,\frac{\sqrt{2}}{X^2}\iota_a\}$.}
\be\label{notsh}
o\mapsto \sqrt{\frac{2}{-X^2}}{\bar \iota},~~~\iota\mapsto \sqrt{\frac{-X^2}{2}}\bar{o},~~~\varphi^\Delta\mapsto (-X^2)^{\Delta-1}\varphi^{\Delta}\,,
\ee
which is not to be confused with the shadow transform.
This takes us from table~\ref{tab:CPWspinors} to table~\ref{tab:SHCPWspinors}. 
\begin{table}[ht]
\renewcommand*{\arraystretch}{1.3}
    \centering
    \begin{tabular}{l|l|c}
$\widetilde{\Phi}_{\Delta,J}$&Wavefunction& Reference\\
\hline
 $\widetilde{\Phi}_{\Delta,0}$ & $\tvarphi^{\Delta}=(-X^2)^{\Delta-1}\varphi^{\Delta}$& \cite{Pasterski:2017kqt}\\
 $\widetilde{\Phi}_{\Delta,+\frac{1}{2}}$ & $\widetilde{{\overline\psi}}^{\Delta}_{J=+\frac{1}{2}}=\sqrt{2}{\bar \iota}(-X^2)^{\Delta-\frac{3}{2}} \varphi^\Delta$ &---\\
  $\widetilde{\Phi}_{\Delta,+1}$ & $\tA_{J=+1}^{\Delta}=o{\bar \iota}(-X^2)^{\Delta-1}\varphi^{\Delta} $& \cite{Pasterski:2017kqt}\\
 $\widetilde{\Phi}_{\Delta,+\frac{3}{2}}$ & $\widetilde{{\overline\chi}}^{\Delta}_{J=+\frac{3}{2}}=\sqrt{2}o {\bar \iota}{\bar \iota}(-X^2)^{\Delta-\frac{3}{2}}\varphi^{\Delta}$&---\\
  $\widetilde{\Phi}_{\Delta,+2}$  &  $\th_{J=+2}^{\Delta,\pm}=o o{\bar \iota}{\bar \iota}(-X^2)^{\Delta-1}\varphi^{\Delta}$ &\cite{Pasterski:2017kqt}
 \end{tabular}
 \caption{Negative helicity conformal shadow primary wavefunctions in spinor notation.}
 \label{tab:SHCPWspinors}
\end{table}
We will see below that the factor of $\sqrt{2}$ in the map matches the normalization of the conformal primaries and their shadows for the spin-1 and spin-2 field strengths.

We conclude this section by observing that this tabulation of conformal primary wavefunctions adds another conformal Goldstone mode to the mix.  While the large gauge symmetry of the gravitino\be\label{largegauge}
\chi\mapsto \chi+\p_\mu \psi
\ee was studied in connection to a corresponding soft theorem in~\cite{Lysov:2015jrs,Avery:2015iix}, and that soft theorem was connected to a conformal soft theorem for the Mellin representatives in~\cite{Fotopoulos:2020bqj}, we are pleased to identify the spin-$\frac{3}{2}$ conformal primary Goldstino in this paper.

Although the large gauge symmetry of the gravitino~\eqref{largegauge} is eliminated by our gauge fixing  condition~\eqref{gaugefix}, we will see that for certain values of $\Delta$ the spin-$\frac{3}{2}$ conformal primary reduces to being pure gauge. To this end we introduce the analog of a `field strength' 
\be
{\cal f}_{\mu\nu}\equiv\p_\mu\chi_{\nu}-\p_\nu\chi_{\mu}.
\ee
This combination will vanish precisely when the conformal primary is pure gauge
\be
{\cal f}_{\mu\nu}=0\Leftrightarrow\chi_\mu=\p_\mu \psi\,.
\ee
Evaluating this expression allows one to read off the entries in tables~\ref{table:Goldstone} and~\ref{table:shadowGoldstone}.

For the non-shadow modes, we will be more explicit.  Note that the Mellin representative of~\cite{Fotopoulos:2020bqj} in table~\ref{tab:MellinReps} and the conformal primary wavefunction in table~\ref{tab:CPWspinors} differ by a large gauge transformation of the type~\eqref{largegauge}, up to a normalization. For the positive helicity spin-$\frac{3}{2}$ conformal primary~\eqref{nonshadowLeft}, we have
\be\label{melvcpw}
\chi^{\Delta,\pm}_{\mu;J=+\frac{3}{2}}=\frac{\Delta-\frac{1}{2}}{\Delta+\frac{1}{2}}\frac{(\pm i)^\Delta}{\Gamma(\Delta)}\epsilon_{\mu;+}o^\pm\phi^{\Delta,\pm}+\p_\mu \beta^{\Delta,\pm}_+\,
\ee
where
\be
\beta^{\Delta,\pm}_+=\Delta o^\pm\alpha^{\Delta,\pm}_+\,, \quad \quad \alpha^{\Delta,\pm}_+=\frac{\epsilon_+\cdot X_\pm}{\Delta(-q\cdot X_\pm)^\Delta}\,,
\ee
and $\alpha^{\Delta,\pm}_J$ is the residual gauge term for spin-1 seen in~\cite{Donnay:2020guq}. 
The conformal primary~\eqref{melvcpw} reduces to pure gauge precisely for $\Delta=\frac{1}{2}$, matching the conformally soft pole for the celestial amplitudes identified in~\cite{Fotopoulos:2020bqj}.  In particular, we see from~\eqref{melvcpw} that
\be
\chi_{\mu;J=+\frac{3}{2}}^{\frac{1}{2},\pm}
=\partial_\mu \beta_+^{\frac{1}{2},\pm}.
\ee
One obtains a similar expression for the $\Delta=\frac{1}{2}$ negative helicity spin-$\frac{3}{2}$ conformal primary~\eqref{nonshadowRight}. 
Table~\ref{table:Goldstone} summarizes this result, adding the spin-$\frac{3}{2}$ Goldstino to the spin-1 and spin-2~\cite{Cheung:2016iub,Pasterski:2017kqt} conformal primary Goldstone modes of spontaneously broken asymptotic symmetries.
Meanwhile, the spin-$\frac{3}{2}$ conformal shadow primaries~\eqref{shadowLeft}-\eqref{shadowRight} and their opposite helicity expressions become pure gauge for $\Delta=\frac{3}{2}$. This is summarized in table~\ref{table:shadowGoldstone}, whose entries correspond to the shadow transforms of the conformal primaries appearing in table~\ref{table:Goldstone}.

Curiously, whenever the shadow modes for spin $1\le s \le2$ are pure gauge, they have conformal dimension matching the classical scaling dimension for the corresponding free field. Moreover, away from isolated points where both $X^2=0$ and $q\cdot X=0$, conformal shadow primaries of any spin $s=\{0,\frac{1}{2},1,\frac{3}{2},2\}$ obey the standard radiative fall-offs. Indeed, it is these shadow modes that pick out the stress tensor~\cite{Kapec:2016jld}, supercurrent~\cite{Fotopoulos:2020bqj}, and large gauge current~\cite{Donnay:2018neh} in amplitudes.  This is done via charges constructed from the shadow Goldstone and Goldstino modes~\cite{Donnay:2020guq,PPP}. 
\begin{table}[ht]
\renewcommand*{\arraystretch}{1.3}
\centering
\begin{tabular}{|c|c|c|cc|}
\hline
  & \multicolumn{1}{c|}{${A}^{\Delta}_{\mu}$} & \multicolumn{1}{c|}{$\chi^{\Delta}_{\mu}$}  & \multicolumn{2}{c|}{${h}^{\Delta}_{\mu\nu}$}\\
  \hline
 $\Delta$ &1 &$\frac{1}{2}$ &  1&  0 \\
 symmetry & large $U(1)$ & local SUSY  & supertranslation & shadow superrotation $\in$ Diff$(S^2)$\\
\hline
\end{tabular}
\caption{Goldstone modes of spontaneously broken asymptotic symmetries for particles with spin $1\le s\le2$.  
}
 \label{table:Goldstone}
\end{table}
\begin{table}[ht]
\renewcommand*{\arraystretch}{1.3}
\centering
\begin{tabular}{|c|c|c|cc|}
\hline
  & \multicolumn{1}{c|}{${\tA}^{\Delta}_{\mu}$} & \multicolumn{1}{c|}{$\tilde{\chi}^{\Delta}_{\mu}$}  & \multicolumn{2}{c|}{${\th}^{\Delta}_{\mu\nu}$}\\
  \hline
 $\Delta$ &1 &$\frac{3}{2}$ &  1&  2 \\
 symmetry & large $U(1)$ & local SUSY  & supertranslation & superrotation\\
\hline
\end{tabular}
\caption{Shadow Goldstone modes of spontaneously broken asymptotic symmetries for particles with spin $1\le s\le2$.  
}
 \label{table:shadowGoldstone}
\end{table}

In equations, using the compact notation $\Phi_{\Delta,J}$ for the conformal primary wavefunctions and a similar notation ${\cal O}_{\Delta,J}$ for the quantum fields, we have
\be\begin{array}{rl}
{\cal O}_{|J|}=\sum\limits_{J=\pm|J|}\int d^2 w \int_{1-i\infty}^{1+i\infty}(-id\Delta)& \Big[{\cal N}^+_{2-\Delta,|J|}\Phi_{2-\Delta,-J}(X_+;w,\bw)a_{\Delta,J}(w;\bw)\\
&~+{\cal N}^-_{\Delta,|J|}\Phi_{\Delta,J}(X_-;w,\bw)a_{\Delta,J}(w;\bw)^\dagger\Big]\,,
\end{array}\ee
where, as in~\cite{Donnay:2020guq}, the normalization factors are designed to guarantee the canonical commutation relations
\be
[a_{\Delta,J}(w;\bw),a_{\Delta',J'}(w';\bw')]=\delta_{JJ'}\delta^{(2)}(w-w')\boldsymbol{\delta}(i(\Delta+\Delta'^*-2))\,,
\ee
and follow from the inner product of the conformal primary wavefunctions.  These factors were computed for the spin-1 $({\cal O}_{1}=A)$ and spin-2 $({\cal O}_{2}=h)$ cases in~\cite{Donnay:2020guq}.  We will compute the remaining ones (scalar and fermionic cases) in a companion paper~\cite{PPP}.

From these mode expansions, we can define the operators
\be
Q_{\Delta,J}(w,\bw)\equiv i({\cal O}_{|J|},\Phi^+_{\Delta^*,-J}(X;w,\bw))\,,
\ee
with similar expressions for the shadow modes.  For any $(\Delta,J)$ for which the field appearing in this inner product is a Goldstone mode, this operator reduces to the canonical charge for that asymptotic symmetry.\footnote{This may require an appropriate renormalization as in~\cite{Donnay:2020guq}.}  In~\cite{PPP}, we will detail the appropriate inner products for the spin-$\frac{3}{2}$ missing in~\cite{Donnay:2020guq}.  We will also show that there exists a corresponding interpretation for spin-0 and spin-$\frac{1}{2}$ operators for $\Delta=1$.  While there is no corresponding gauge symmetry, these operators are related by supersymmetry to fields which do possess large gauge freedom.

\section{Incrementing Spin with Operators}\label{Mellinspin}

In section~\ref{susyq}, we recall the action of the generators of the (super-)Poincar\'e algebra on bosonic and fermionic Mellin representatives studied in~\cite{Stieberger:2018onx,Fotopoulos:2020bqj} and extend this formalism to allow us to make an arbitrary number of half-integer steps in spin between our conformal primary wavefunctions. We then phrase the classical double copy relation between the curvatures of the bosonic Mellin representatives as an celestial operator statement in section~\ref{sec:CDCops}. Finally, we contrast these Mellin expressions with the corresponding ones for conformal primary wavefunctions and show in section~\ref{gaugequiv} that the Lie derivative relating perturbative Mellin representatives to conformal primary wavefunctions exponentiates.

\subsection{Shifting Spin with Supersymmetry}\label{susyq}
We will now discuss operators that shift the spin of conformal primaries by half integer steps. The natural physical setting for this would involve supersymmetry. The authors of~\cite{Fotopoulos:2020bqj} studied the $\mathcal{N}=1$ supersymmetric extension of the 4D BMS algebra. Supermultiplets of conformal primary wave functions for bosonic and fermionic states are related via supersymmetry.  In 2D superspace, the supersymmetry generators are given by\footnote{Note that our definitions differ from~\cite{Fotopoulos:2020bqj} since we continue to use the conventions of~\cite{Elvang:2013cua}.}
\be\label{susyQs}
{\Q}_{a}=\p_\theta|q]_{a}e^{\p_\Delta/2}\,,\quad{\bar \Q}_{\dot a}=\theta\langle q|_{\dot a}e^{\p_\Delta/2}
\ee
where $\theta$ is a Grassmann variable. These obey the expected 4-dimensional supersymmetry algebra 
\be
\{\Q_a,\bar{\Q}_{\dot a}\}=-\sigma^\mu_{a{\dot a}}{\cal P}_\mu\,,
\ee
where the translation generator of the Poincar\'e algebra is given by the differential operator~\cite{Stieberger:2018onx}
\be\label{calP}
{\cal P}_\mu=q_\mu e^{\p_\Delta}\,.
\ee
The action of~\eqref{calP} on an operator or a wavefunction
\be\label{peq}
\mathcal{P}^\mu\mathcal{O}_{\Delta,J}(w,\bw)=q^\mu \mathcal{O}_{\Delta+1,J}(w,\bw)\,,
\ee
shifts its $SL(2,\mathbb{C})$ conformal weight by an amount opposite to that of $q^\mu$~\eqref{qtra}, and therefore matches our expectation that a Lorentz vector operator has conformal dimension $\Delta=0$.
Because the null vectors~\eqref{eq:pol} and~\eqref{tetrad}, as well as the spinors~\eqref{spinframespinors}, have no $\Delta$-dependence, it suffices to consider~\eqref{peq} for the spin-0 wavefunction
\begin{equation}\label{Pphi}
 \mathcal{P}^\mu \varphi^\Delta = q^\mu \varphi^{\Delta+1}\,,
\end{equation}
to infer the action of~\eqref{calP} on all spin $s=\{0,\frac{1}{2},1,\frac{3}{2},2\}$ Mellin representatives and conformal primary wavefunctions (which agree for spin-0 and spin-$\frac{1}{2}$ up to normalization). In celestial amplitudes, the translation generator relates operator insertions of different conformal dimensions.

The fermionic analogue of~\eqref{Pphi} is given by the action of the differential operator in the supersymmetry generators~\eqref{susyQs} relating the spin-$\frac{1}{2}$ to the spin-0 wavefunction
\be\label{ddel}
\psi^\Delta_{J=+\frac{1}{2}}=|q]e^{\p_\Delta/2}\varphi^\Delta=|q]\varphi^{\Delta+\frac{1}{2}}\,.
\ee
This action is equivalent to the relation displayed in table~\ref{tab:CPWspinors} via spin frame multiplication and shows that, similar to the action of the translation generator in celestial amplitudes, the supercharge relates operator insertions of different conformal dimensions.

By dropping the $\theta$ dependence from the supercharges of~\cite{Fotopoulos:2020bqj} -- which made them satisfy the ${\cal N}=1$ supersymmetry algebra but also makes them nilpotent -- we can define a more general spin-shifting operator that takes us between conformal primary wavefunctions of arbitrary spin. Let us start by defining
\be
T_{a}^{~\dot a}\equiv\frac{\sqrt{2}}{X^2}X_{a\dot{b}} \varepsilon^{\dot b\dot a}
,~~~
{\bar T}_{\dot a}^{~a}\equiv\frac{-\sqrt{2}}{X^2}\varepsilon_{\dot{a} \dot{b}}\bar{X}^{\dot{b} a},
\ee
where $\bar{X}^{\dot{b} a}\equiv X_\mu(\bar{\sigma}^\mu)^{\dot{b} a}$, so that
\be\label{eq:oio}
{o}_{a}=T_{ a}^{~\dot{a}}\bar{\iota}_{\dot a},~~~\bar{o}_{\dot a}={\bar T}_{\dot a}^{~a}\iota_a.
\ee
It is natural to define a field with all left or right handed indices, depending on the sign of $J$.  For $J>0$, we let
\be
\hat{\Phi}_{\Delta,J;a_1...a_{2J}}=\prod_{i=1}^{\lfloor J\rfloor}T_{a_{\lceil J\rceil+i}}^{~{\dot a}_i}\Phi_{\Delta,J;a_1...a_{\lceil J\rceil}\dot{a}_{1}...\dot{a}_{\lfloor J\rfloor}}.
\ee
It is straightforward to check from~\eqref{eq:oio} and table~\ref{tab:CPWspinors} that these modes obey
\be
\hat{\Phi}_{\Delta,J}=(o)^{2J}\hat{\Phi}_{\Delta,0}=\Big(|q]e^{\p_{\Delta}/2}\Big)^{2J}\hat{\Phi}_{\Delta,0}=|q]^{2J}\hat{\Phi}_{\Delta+J,0}.
\ee
For $J<0$, we use $\bar{T}$ to write
\be
\hat{\Phi}_{\Delta,J;{\dot a}_1...{\dot a}_{2|J|}}=\prod_{i=1}^{\lfloor -J\rfloor}\bar{T}_{\dot{a}_{\lceil -J\rceil +i}}^{~a_i}\Phi_{\Delta,J;{\dot a}_1...{\dot a}_{\lceil -J\rceil}{a}_{1}...{a}_{\lfloor -J\rfloor}}
\ee
so that 
\be
\hat{\Phi}_{\Delta,J}=(\bar{o})^{2|J|}\hat{\Phi}_{\Delta,0}=\Big(\langle q|e^{\p_{\Delta}/2}\Big)^{2|J|}\hat{\Phi}_{\Delta,0}=\langle q|^{2|J|}\hat{\Phi}_{\Delta+|J|,0}.
\ee
We thus have a multiplet of fields $\hat{\Phi}_{\Delta,J}$ constructed from radiative solutions of the equations of motion ${\Phi}_{\Delta,J}$.  A key point is that the map $\Phi_{\Delta,J}\mapsto \hat{\Phi}_{\Delta,J}$ can be inverted to return a radiative solution with the corresponding spin. Since the fields $\hat{\Phi}_{\Delta,J}$ are totally symmetric in their spinor index, we can use
\be
({ T}^{-1})_{\dot{ a}}^{~{a}}=-\frac{1}{\sqrt{2}}\varepsilon_{\dot{a}\dot{b}}\bar{X}^{\dot{b} a},~~~({\bar T}^{-1})_a^{~\dot{a}}=\frac{1}{\sqrt{2}}X_{a\dot{b}} \varepsilon^{\dot{b}\dot{a}}
\ee
to write
\be
\Phi_{\Delta,J;a_1...a_{\lceil J\rceil}\dot{a}_{1}...\dot{a}_{\lfloor J\rfloor}}
=\prod_{i=1}^{\lfloor J\rfloor} ({ T}^{-1})_{\dot{a}_{i}}^{~a_{\lceil J\rceil+i}}\hat{\Phi}_{\Delta,J;a_1...a_{2J}}
\ee
for $J>0$, while for $J<0$ we have
\be
\Phi_{\Delta,J;{\dot a}_1...{\dot a}_{\lceil -J\rceil}{a}_{1}...{a}_{\lfloor -J\rfloor}}
=\prod_{i=1}^{\lfloor -J\rfloor}({ \bar T}^{-1})_{{a}_{i}}^{~{\dot a}_{\lceil -J\rceil+i}}\hat{\Phi}_{\Delta,J;\dot{a}_1...\dot{a}_{2J}}.
\ee
We can combine these results to write a differential operator that shifts us between conformal primary wavefunctions $\Phi_{\Delta,J}$ of different spin, avoiding the new multiplet $\hat{\Phi}_{\Delta,J}$.  For $J>0$, we have
\be
\Phi_{\Delta,J;a_1...a_{\lceil J\rceil}\dot{a}_{1}...\dot{a}_{\lfloor J\rfloor}}
=\left[\prod_{i=1}^{\lceil J\rceil} |q]_{{a}_{i}}e^{\p_\Delta/2}\prod_{j=1}^{\lfloor J\rfloor} ({ T}^{-1}|q])_{\dot{a}_{j}}e^{\p_\Delta/2}\right]{\Phi}_{\Delta,0}
\ee
and for $J<0$ 
\be
\Phi_{\Delta,J;{\dot a}_1...{\dot a}_{\lceil -J\rceil}{a}_{1}...{a}_{\lfloor -J\rfloor}}
=\left[\prod_{i=1}^{\lceil -J\rceil} \langle q|_{{\dot a}_{i}}e^{\p_\Delta/2}\prod_{j=1}^{\lfloor -J\rfloor} ({ \bar{T}}^{-1}\langle q|)_{{a}_{j}}e^{\p_\Delta/2}\right]{\Phi}_{\Delta,0}.
\ee
If we were interested in amplitudes instead of wavefunctions, we would expect to replace every appearance of $X^\mu$ with a differential operator.  In the plane wave basis this would be $i\p_{k^\mu}$, whereas for conformal primary wavefunctions we would use
\be\label{xop}
\p_{q^\mu}\varphi^{\Delta}=\Delta X_\mu \varphi^{\Delta+1}~~\Rightarrow X_\mu \mathcal{O}_\Delta= \frac{1}{\Delta-1} e^{-\p_\Delta}\p_{q^\mu}\mathcal{O}_\Delta.
\ee
We leave subtleties regarding on-shell restrictions that complicate such manipulations in amplitudes to future work.  It is worth pointing out that the $\Delta$-dependence of~\eqref{xop} implies that the operator ordering matters in this representation.

We can also use the operators defined in this section to create a multiplet for the shadow modes. Observing that
\be
T^{-1}=\frac{X^2}{2}\bar{T},~~~\bar{T}^{-1}=\frac{X^2}{2}T
\ee
and letting $S$ denote the action of the map~\eqref{notsh} on the spinors $\{o,\iota\}$, we find the intertwining relation
\be
ST=-\bar{T}S,~~S^2=1
\ee
meaning that up to the price of an overall sign, we can apply the map $S$ before or after mapping to a multiplet with uniform indices.  

Finally, we note that one can also go back down in spin.  Using $\iota_a o^a=\bar{\iota}_{\dot a} \bar{o}^{\dot a}=\sqrt{2}$,
we have 
\be
\hat{\Phi}_{\Delta,J;a_{1}...a_{2J}}=o_{a_1}\hat{\Phi}_{\Delta,J-\frac{1}{2};a_2...a_{2J}},~~~\hat{\Phi}_{\Delta,J-\frac{1}{2};a_2...a_{2J}}=\frac{1}{\sqrt{2}}(\bar{T}^{-1}\bar{o})^{a_1}\hat{\Phi}_{\Delta,J;a_1...a_{2J}}
\ee
for $J>0$ or, equivalently, for the original radiative modes\footnote{Note that depending on the use case, these expressions can be written in two different suggestive forms.  First, as a relation between shifted $\Delta$ and $J$; second, as a differential operator shifting $\bar{h}=\frac{1}{2}(\Delta-J)$.  For example:
\be
\Phi_{\Delta,J;a_1...a_{\lceil J\rceil}\dot{a}_{1}...\dot{a}_{\lfloor J\rfloor}}=|q]_{a_1}\Phi_{\Delta+\frac{1}{2},J-\frac{1}{2};a_2...a_{\lceil J\rceil}\dot{a}_{1}...\dot{a}_{\lfloor J\rfloor}}\Rightarrow|q]e^{\p_{\bar h}/2}\Phi_{\Delta,J}.\notag
\ee
The final form is schematic on wavefunctions with spinor indices but could be interesting from the point of operators in the 2D celestial CFT in contexts where one considers analytically continuing $J$~\cite{Caron-Huot:2017vep}.}
\be
\badat{2}
\Phi_{\Delta,J;a_1...a_{\lceil J\rceil}\dot{a}_{1}...\dot{a}_{\lfloor J\rfloor}}&=~|q]_{a_1}e^{\p_\Delta/2}\Phi_{\Delta,J-\frac{1}{2};a_2...a_{\lceil J\rceil}\dot{a}_{1}...\dot{a}_{\lfloor J\rfloor}}\\
\Phi_{\Delta,J-\frac{1}{2};a_2...a_{\lceil J\rceil}\dot{a}_{1}...\dot{a}_{\lfloor J\rfloor}}&=~\frac{1}{\sqrt{2}}(\bar{T}^{-1}\langle q|)^{a_1}e^{\p_\Delta/2}\Phi_{\Delta,J;a_1...a_{\lceil J\rceil}\dot{a}_{1}...\dot{a}_{\lfloor J\rfloor}}
\eadat
\ee
for $J>0$ and odd, and 
\be
\badat{2}
\Phi_{\Delta,J;a_1...a_{\lceil J\rceil}\dot{a}_{1}...\dot{a}_{\lfloor J\rfloor}}&=~(T^{-1}|q])_{\dot{a}_1}e^{\p_\Delta/2}\Phi_{\Delta,J-\frac{1}{2};a_1...a_{\lceil J\rceil}\dot{a}_{2}...\dot{a}_{\lfloor J\rfloor}}\\
\Phi_{\Delta,J-\frac{1}{2};a_1...a_{\lceil J\rceil}\dot{a}_{2}...\dot{a}_{\lfloor J\rfloor}}&=~\frac{-1}{\sqrt{2}}\langle q|^{\dot{a}_1}e^{\p_\Delta/2}\Phi_{\Delta,J;a_1...a_{\lceil J\rceil}\dot{a}_{1}...\dot{a}_{\lfloor J\rfloor}}
\eadat
\ee
for $J>0$ and even. It is straightforward to write down the analogues for negative helicity. 

We have thus found that incrementing ($+$ decrementing) spin in wavefunctions by half-integer steps is achieved by a modified ($\theta$-stripped) version of the supersymmetry generators which, in turn, can be seen as the ``square root'' of the translation generator. Integer steps in spin can be achieved with even powers of these operators within the multiplet $\hat{\Phi}_{\Delta,J}$, via multiplication by $\epsilon_J\sim \partial_J q$ for Mellin representatives, or by $m$ or $\bar{m}$ for conformal primary wavefunctions ${\Phi}_{\Delta,J}$.  The latter amounts to the classical double copy which we will discuss in detail in sections~\ref{sec:CDCWeyl} and~\ref{sec:CDCKS}. The perturbative double copy for amplitudes was recently shown to hold for celestial amplitudes in the form of an operator statement that involves the action of~\eqref{calP} on individual external states given by the Mellin representatives~\cite{Casali:2020vuy}. We will now demonstrate its classical analogue involving the curvatures for the Mellin representatives.

\subsection{Celestial Operator Double Copy}\label{sec:CDCops}

The perturbative double copy for celestial amplitudes involving (the Mellin transforms of) plane waves has a simple classical analogue. 
Electromagnetic waves $\epsilon_{\mu;\ell} e^{ik\cdot X}$ in gauge theory with field strength $
 F_{\mu\nu;\ell}=i(k_\mu\epsilon_{\nu;\ell}-k_\nu \epsilon_{\mu;\ell})e^{i k\cdot X}$
and gravity waves $\epsilon_{\mu;\ell}\epsilon_{\nu;\ell} e^{ik\cdot X}$
with Riemann tensor 
$ R_{\mu\nu\rho\sigma;\ell}=\frac{1}{2}(k_\mu\epsilon_{\nu;\ell}-k_\nu \epsilon_{\mu;\ell})(k_\rho\epsilon_{\sigma;\ell}-k_\sigma \epsilon_{\rho;\ell})e^{i k\cdot X}$ 
satisfy an obvious squaring relation after stripping off the scalar wavefunction $e^{i k\cdot X}$.
This simple relation between the curvatures gets promoted to an operator valued double copy in the conformal basis.

The curvatures of the integer Mellin representatives of table~\ref{tab:MellinReps} are easily computed in terms of the {\it celestial momenta} $K^\Delta_\mu$ introduced in~\cite{Casali:2020vuy} by
\begin{equation}\label{celestialK}
   \partial_\mu \phi^\Delta = \frac{\Delta q_\mu}{-q\cdot X}\phi^\Delta\equiv K^\Delta_\mu \phi^\Delta\,.
\end{equation}
The field strength for the spin-one Mellin representative $\epsilon_{\mu;J} \phi^\Delta$ is
\begin{equation}\label{FieldstrengthV}
 F^{\Delta}_{\mu\nu;J}=(K^\Delta_\mu \epsilon_{\nu;J}-K^\Delta_\nu \epsilon_{\mu;J})\phi^{\Delta}\,,
\end{equation}
and the Riemann tensor for the the flat background perturbed by the spin-two Mellin representative $g^\Delta_{\mu\nu;J}=\eta_{\mu\nu}+\epsilon_{\mu;J} \epsilon_{\nu;J}  \phi^\Delta$ with inverse $g^{\Delta\, \mu\nu}_{J}=\eta^{\mu\nu}-\epsilon^\mu_J \epsilon^\nu_J \phi^\Delta$
is
\begin{equation}\label{RiemannV}
 R^{\Delta}_{\mu\nu\rho\sigma;J}=-\frac{1}{2}\Big(1+\frac{1}{\Delta}\Big) (K^\Delta_\mu \epsilon_{\nu;J}-K^\Delta_\nu \epsilon_{\mu;J}) (K^\Delta_\rho \epsilon_{\sigma;J}-K^\Delta_\sigma \epsilon_{\rho;J})\phi^{\Delta}\,.
\end{equation}
The $\Delta$ dependent factor is familiar from the double copy for celestial amplitudes~\cite{Casali:2020vuy} and arises from the space-time dependence of the celestial momenta. Expressing the latter via the action of the translation generator~\eqref{calP} on the scalar wavefunction
\begin{equation}
    {\cal P}_\mu \phi^\Delta = K^\Delta_\mu \phi^\Delta\,, \quad  {\cal P}_\mu {\cal P}_\nu \phi^\Delta = \Big(1+\frac{1}{\Delta}\Big) K^\Delta_\mu K^\Delta_\nu \phi^\Delta\,,
\end{equation}
promotes the celestial momenta in~\eqref{FieldstrengthV} and \eqref{RiemannV} to momentum operators. Defining
\begin{equation}
    F^{\Delta}_{\mu\nu;J}=\mathcal{F}_{\mu\nu;J}\phi^\Delta\,, \quad R^{\Delta}_{\mu\nu\rho\sigma;J}=\mathcal{R}_{\mu\nu\rho\sigma;J}\phi^\Delta\,,
\end{equation}
we see that the Riemann tensor is related to the field strength via the operator valued squaring relation 
\begin{equation}
 -2\mathcal{R}_{\mu\nu\rho\sigma;J}=\mathcal{F}_{\mu\nu;J}\mathcal{F}_{\rho\sigma;J}\,,
\end{equation}
when acting on the conformal wavefunction $\phi^\Delta$.
This resonates with the double copy for celestial amplitudes~\cite{Casali:2020vuy}, where the operator valued numerator in gauge theory squares to the operator valued gravity numerator when acting on the amplitude of scalar wavefunctions. Indeed, they have the same origin.

\subsection{Gauge Equivalence for Finite Deformations}\label{gaugequiv}

While the Mellin representatives $\epsilon_{\mu;J}\epsilon_{\nu;J} \phi^\Delta$
and the corresponding conformal primary wavefunctions of $h^\Delta_{\mu\nu;J}=m_{\mu;J} m_{\nu;J}\varphi^\Delta$
were originally obtained as solutions to the linearized equations of motion~\eqref{lineom}, i.e. as perturbations around the Minkowski background, we will show in section~\ref{sec:CDCKS} that they are actually exact solutions to the non-linear Einstein equations, by virtue of the Kerr-Schild double copy. In this section, we show that the Lie derivative along the diffeomorphism relating these wavefunction perturbations exponentiates, and thereby demonstrate that the non-linear solutions constructed from the Mellin representatives and the conformal primary wavefunctions are gauge equivalent.

In~\cite{Donnay:2020guq} it was pointed out that the Mellin representative and the conformal primary wavefunction
\begin{equation}\label{hDelta}
 h^{\Delta,\pm}_{\mu\nu;J}(X^\mu;w,\bw)
 =\frac{\Delta-1}{\Delta+1}\frac{(\pm i)^\Delta}{\Gamma(\Delta)} \epsilon_{\mu;J}\epsilon_{\nu;J}\phi^{\Delta,\pm}+\p_\mu \zeta^{\Delta,\pm}_{\nu;J}+\p_\nu \zeta^{\Delta,\pm}_{\mu;J}\,,
\end{equation}
are related, up to a normalization, by a diffeomorphism
\begin{equation}\label{zetaDiff}
\zeta^{\Delta,\pm}_{\mu;J}=\frac{1}{2(\Delta+1)}\left(\frac{\epsilon_{\mu;J} (\epsilon_{J}\cdot X_\pm)}{(-q\cdot X_\pm)^\Delta}+\frac{1}{2}\frac{q_\mu (\epsilon_{J}\cdot X_\pm)^2}{(-q\cdot X_\pm)^{\Delta+1}}\right)\,.
\end{equation}
In the above expression, the Lie derivative is acting on the flat background around which both wavefunctions are perturbations. It is straightforward to show that 
\be\begin{array}{ll}\label{lie2}
\mathcal{L}_{\beta \zeta^\Delta_J}\Big(\eta+\alpha \epsilon_{J}\epsilon_{J}\phi^{\Delta,\pm}\Big)_{\mu\nu}&=\alpha\beta  \Delta\epsilon_{\mu;J}\epsilon_{\nu;J}(\zeta^\Delta_{J}\cdot l) \phi^\Delta+\beta (\p_\mu \zeta^{\Delta}_{\nu;J}+\p_\nu \zeta^{\Delta}_{\mu;J})\,\\
&~~+\alpha\beta (\p_\mu \zeta^{\Delta;\sigma}_{J}\epsilon_{\nu;J}+\p_\nu \zeta^{\Delta;\sigma}_J\epsilon_{\mu;J})\epsilon_{J;\sigma}\phi^\Delta\\
&=\beta(\p_\mu \zeta^{\Delta}_{\nu;J}+\p_\nu \zeta^{\Delta}_{\mu;J})\,\\
\end{array}\ee
where we have used~\eqref{qepsIDs}, the spacetime independence of the null vector $q$, as well as
\begin{equation}\label{celestialKl}
    K^\Delta_\mu =\Delta l_\mu\,.
\end{equation}
Since $\zeta^\Delta_J$ only involves the span of $\epsilon_J$ and $q$, we can raise and lower indices with the flat metric. 
The final form of~\eqref{lie2} demonstrates that we can exponentiate the Lie derivative and that the exact gravitational solutions constructed from the Mellin representatives and conformal primaries are gauge equivalent.

To complete the discussion, we report here the curvatures of the conformal primary wavefunctions $A^\Delta_{\mu;J}=m_{\mu;J} \varphi^\Delta$ and $h^\Delta_{\mu\nu;J}=m_{\mu;J} m_{\nu;J}\varphi^\Delta$ of table~\ref{tab:CPWspinors}. The field strength for the $J=+1$ conformal primary is given by
\begin{equation}\label{KSFieldStrength0}
F^\Delta_{\mu\nu;J=+1}=(\Delta-1)\Big(l_\mu m_\nu-l_\nu m_\mu\Big) \varphi^\Delta
\,,
\end{equation}
while the Riemann tensor for the $J=+2$ conformal primary is
\begin{equation}\label{KSRiemann0}
 R^\Delta_{\mu\nu \rho\sigma;J=+2}
  =-\frac{1}{2} \Delta(\Delta-1)(l_\mu m_\nu-l_\nu m_\mu) (l_\rho m_\sigma-l_\sigma m_\rho) \varphi^\Delta
\,,
\end{equation}
and equivalent expressions for $J\mapsto-J$ with $m\mapsto\bar{m}$. Here, we have chosen to write these field strengths in terms of our tetrad~\eqref{tetrad}, due to its nice conformal covariance properties~\eqref{tetradcov}. Comparing~\eqref{FieldstrengthV}-\eqref{RiemannV} to~\eqref{KSFieldStrength0}-\eqref{KSRiemann0} using~\eqref{celestialKl}, we observe that the curvatures involving the Mellin representatives and the conformal primary wavefunctions are the same up to an overall normalization (which will be an important distinction in the conformally soft limit associated to asymptotic symmetries below), in line with the result~\eqref{lie2} above.

For bookkeeping's sake, let us also write the curvatures for the conformal shadow primaries of Table~\ref{tab:SHCPWspinors}: 
\begin{equation}\label{SHKSFieldStrength0}
\tF^{\Delta}_{\mu\nu;J=+1}=(\Delta-1)\Big(n_\mu m_\nu-n_\nu m_\mu\Big) \frac{2}{X^2}\tvarphi^{\Delta}
\,,
\end{equation}
and 
\begin{equation}\label{SHKSRiemann0}
 \tR^{\Delta}_{\mu\nu \rho\sigma;J=+2}
  =-\frac{1}{2} (\Delta-1)(\Delta-2)(n_\mu m_\nu-n_\nu m_\mu) (n_\rho m_\sigma-n_\sigma m_\rho) \left(\frac{2}{X^2}\right)^2\tvarphi^{\Delta}.
\,
\end{equation}

\section{Weyl Double Copy for Conformal Primaries}\label{sec:CDCWeyl}

Conformal primary solutions exhibit definite (anti-)self duality -- a property that proves useful when considering scattering in non-trivial backgrounds.\footnote{See, for instance, reference~\cite{Mason:2008jy} and related work for an example application of anti-self dual backgrounds to MHV scattering amplitudes computed in the twistor formalism.} Conformal primaries thus naturally satisfy a Weyl double copy which we now demonstrate.

In vacuum spacetimes, the Riemann tensor coincides with the Weyl tensor $R_{\mu\nu\rho\sigma}=W_{\mu\nu\rho\sigma}$.
The spinorial form of the Weyl tensor is
\begin{equation}
    W_{a\dot{a}b\dot{b}c\dot{c}d\dot{d}}=C_{abcd}\varepsilon_{\dot{a}\dot{b}}\varepsilon_{\dot{c}\dot{d}} +\bar{C}_{\dot{a}\dot{b}\dot{c}\dot{d}}\varepsilon_{ab}\varepsilon_{cd} \,,
\end{equation}
where $C_{abcd}$ and $\bar{C}_{\dot{a}\dot{b}\dot{c}\dot{d}}$ are the anti-self dual and self dual parts of the curvature.  They are completely symmetric and $C_{abcd}=(\bar{C}_{\dot{a}\dot{b}\dot{c}\dot{d}})^*$ if the Lorentzian spacetime is real.
Analogously, the field strength $F_{\mu\nu}$ in gauge theory can be written in spinorial form as
\begin{equation}
    F_{a\dot{a}b\dot{b}}=f_{ab}\varepsilon_{\dot{a}\dot{b}} + f_{\dot{a}\dot{b}}\varepsilon_{ab} \,,
\end{equation}
where $f_{ab}$ and $\bar{f}_{\dot{a}\dot{b}}$ are the anti-self dual and self dual parts of the field strength.  They are symmetric and $f_{ab}=(\bar{f}_{\dot{a}\dot{b}})^*$ if the field strength is real.
Using 
\begin{equation}\label{eq:smn}
    \sigma^{\mu\nu}_{ab}=\sigma^{[\mu}_{a\dot{c}}\bar{\sigma}^{\nu]\dot{c}d} \varepsilon_{db}\,,
\end{equation}
we can express the anti-self dual parts of the curvature and field strength as
\begin{equation}\label{eq:smn2}
    C_{abcd}=\frac{1}{4} W_{\mu\nu\rho\sigma} \sigma^{\mu\nu}_{ab}\sigma^{\rho\sigma}_{cd}\,,
\end{equation}
and
\begin{equation}\label{eq:smn1}
    f_{ab}=\frac{1}{2}F_{\mu\nu}\sigma^{\mu\nu}_{ab}\,.
\end{equation}
The Weyl double copy is defined as~\cite{Luna:2018dpt}
\begin{equation}
    C_{abcd}=\frac{1}{S}f_{(ab}f_{cd)}\,,
\end{equation}
where the scalar $S$ and the field strength spinor $f_{ab}$ are uniquely determined by the Weyl scalar of the gravity solution.

From the decomposition of the tetrad into Weyl spinors~\eqref{spinframe} relation~\eqref{lcr}, we see that the tensor structures appearing in our field strengths~\eqref{KSFieldStrength0} and \eqref{SHKSFieldStrength0} and Weyl tensors~\eqref{KSRiemann0} and \eqref{SHKSRiemann0} reduce to
\begin{equation}
     \badat{4}
     l_{a{\dot a}}m_{b{\dot b}}-l_{b{\dot b}}m_{a{\dot a}}&=\sqrt{2}o_{a}o_{b}\varepsilon_{{\dot a}{\dot b}}\,,\\
     l_{a{\dot a}}{\bar m}_{b{\dot b}}-l_{b{\dot b}}{\bar m}_{a{\dot a}}
 &=\sqrt{2}{\bar o}_{\dot a}{\bar o}_{\dot b}\varepsilon_{ a b}\,,\\
      n_{a{\dot a}}m_{b{\dot b}}-n_{b{\dot b}}m_{a{\dot a}}
 &=-\sqrt{2}{\bar \iota}_{\dot a}{\bar \iota}_{\dot b}\varepsilon_{ a b}\,,\\
 n_{a{\dot a}}{\bar m}_{b{\dot b}}-n_{b{\dot b}}{\bar m}_{a{\dot a}}
 &=-\sqrt{2}{\iota}_{a}{\iota}_{b}\varepsilon_{ {\dot a}{\dot  b}}\,.
\eadat
\end{equation}
We thus notice that 
 \begin{itemize}
 \item the conformal primaries with positive helicity $\{A^\Delta_{J=+1},h^\Delta_{J=+2}\}$ are anti-self dual,
 \item the conformal primaries with negative helicity $\{A^\Delta_{J=-1},h^\Delta_{J=-2}\}$ are self dual,
  \item the conformal shadow primaries with negative helicity $\{\tA^\Delta_{J=+1},\th^\Delta_{J=+2}\}$ are self dual,
   \item the conformal shadow primaries with positive helicity $\{\tA^\Delta_{J=-1},\th^\Delta_{J=-2}\}$ are anti-self dual.
 \end{itemize}
In terms of the Weyl double copy field strengths, we arrive at table~\ref{tab:WeylDC}. Performing a decomposition in terms of the tetrad for the flat metric suffices for analyzing linearized solutions. In section~\ref{sec:CDCKS}, we will show how to generalize this tetrad for the corresponding non-linear solutions. 
\begin{table}[!ht]
\renewcommand*{\arraystretch}{1.3}
    \centering
   \begin{tabular}{c|c|c|c}
&$f_{ab}$& $f_{{\dot a}{\dot b}}$ &$S$\\
\hline
$\Phi_{\Delta,J>0}$&$\sqrt{2}(\Delta-1)o_a o_b\varphi^\Delta$ &$0$ &$-2\frac{\Delta-1}{\Delta}\varphi^\Delta$\\
$\Phi_{\Delta,J<0}$ &$0$ &$\sqrt{2}(\Delta-1)\bar{o}_{\dot{a}} \bar{o}_{\dot b}\varphi^\Delta$ &$-2\frac{\Delta-1}{\Delta}\varphi^\Delta$\\
$\widetilde{\Phi}_{\Delta,J>0}$&$0$ &$-\sqrt{2}(\Delta-1){\bar \iota}_{\dot a}{\bar \iota}_{\dot b}\frac{2}{X^2}\tvarphi^{\Delta}$ &$-2\frac{\Delta-1}{\Delta-2}\frac{2}{X^2}\tvarphi^{\Delta}$\\ $\widetilde{\Phi}_{\Delta,J<0}$&$-\sqrt{2}(\Delta-1)\iota_a\iota_b\frac{2}{X^2}\tvarphi^{\Delta}$ &$0$ &$-2\frac{\Delta-1}{\Delta-2}\frac{2}{X^2}\tvarphi^{\Delta}$
\end{tabular}
\caption{Weyl double copy for conformal primaries.}
\label{tab:WeylDC}
\end{table}

\section{Kerr-Schild Double Copy for Conformal Primaries}\label{sec:CDCKS}

The  Kerr-Schild double copy is a powerful tool for identifying exact solutions of Einstein's equations. It relates a class of Kerr-Schild spacetimes to solutions of Maxwell's equation~\cite{Monteiro:2014cda}.  
Kerr-Schild spacetimes~\cite{Kerr2009} possess the property that they admit coordinates for which the metric $g_{\mu\nu}$ may be written in the form
\begin{equation}
 g_{\mu\nu}=\eta_{\mu\nu}+\KSvec_\mu \KSvec_\nu \varphi\,,
\end{equation}
where $\eta_{\mu\nu}$ is the Minkowski metric, $\varphi$ satisfies the scalar wave equation  $\eta^{\mu\nu}\partial_\mu\partial_\nu\varphi =0$, and the vector $\KSvec_\mu$ has the property that it is null and geodesic with respect to both the Minkowski and the full metric
 \be
 \KSvec^\mu \nabla_\mu \KSvec_\nu\propto \KSvec_\nu ,~~~\KSvec^\mu \p_\mu \KSvec_\nu \propto \KSvec_\nu .
 \ee
Note the inverse metric is simply
\begin{equation}
  g^{\mu\nu}=\eta^{\mu\nu}-\KSvec^\mu \KSvec^\nu \varphi\,,
\end{equation}
and the index on $\KSvec_\mu$ may be raised with either $\eta^{\mu\nu}$ or $g^{\mu\nu}$. The famous property of the Kerr-Schild form is that it linearizes the Ricci tensor with mixed indices
\begin{equation}\label{KSRicci}
 R^{\mu}_{~\nu}=\frac{1}{2} \partial_\lambda \left[\partial^{\mu}(\KSvec^\lambda \KSvec_\nu \varphi)+ \partial_{\nu}(\KSvec^\lambda \KSvec^\mu \varphi)-\partial^\lambda (\KSvec^\mu \KSvec_\nu\varphi)\right]\,, 
\end{equation}
where $\partial^\mu \equiv \eta^{\mu\nu} \partial_\nu$.
The Kerr-Schild double copy now states that if $g_{\mu\nu}$ is a solution to the Einstein equations, then the gauge field given by
\begin{equation}\label{AYM}
 A^{\mathfrak{a}}_\mu=T^{\mathfrak{a}} \KSvec_\mu \varphi\,,
\end{equation}
is a solution to Yang-Mills theory~\cite{Monteiro:2014cda}. 
Since $T^{\mathfrak{a}}$ are just constant color factors here,~\eqref{AYM} lives in a $U(1)$ sector of the gauge theory, with $A_\mu=\KSvec_\mu \varphi$ a solution to Maxwell's equations.

\subsection{Conformal Primaries}\label{sec:nonsh}
It turns out that the spin-one and spin-two conformal primary wavefunctions satisfy a \textit{celestial Kerr-Schild double copy} 
\begin{equation}\label{celestialKerrSchild}
 A^{\Delta}_{\mu;J}=\KSvec_{\mu;J} \varphi^{\Delta}\,, \quad h^{\Delta}_{\mu\nu;J}=\KSvec_{\mu;J}\KSvec_{\nu;J} \varphi^{\Delta}\,,
\end{equation}
with Kerr-Schild vector
\begin{equation}\label{KSvector}
 \KSvec_{\mu;J} =\epsilon_{\mu;J} -q_\mu \frac{\epsilon_J \cdot X}{q\cdot X}\,.
\end{equation}
To see this, notice that the Minkowski background perturbed by the spin-two primary takes the form of a Kerr-Schild metric
\begin{equation}\label{gKerrSchild}
 g^\Delta_{\mu\nu;J}=\eta_{\mu\nu} +\KSvec_{\mu;J} \KSvec_{\nu;J} {\varphi}^{\Delta} \,,
\end{equation}
with inverse
\begin{equation}
 g^{\Delta \,\mu\nu}_{J}=\eta^{\mu\nu} -\KSvec^{\mu}_{\,J} \KSvec^{\nu}_{\,J}{\varphi}^{\Delta} \,.
\end{equation}
With respect to the flat metric $\eta_{\mu\nu}$ and the perturbed metric $g_{\mu\nu}$, the Kerr-Schild vector~\eqref{KSvector} is indeed both null and geodesic
\begin{equation}\label{nullgeodesick}
 g^{\Delta \mu\nu}_{J} \KSvec_{\mu;J} \KSvec_{\nu;J}=\eta^{\mu\nu}\KSvec_{\mu;J} \KSvec_{\nu;J}=0\,,\quad \KSvec^{\mu}_{\,J} \nabla_\mu \KSvec_{\nu;J} = \KSvec^{\mu}_{\,J} \partial_\mu \KSvec_{\nu;J}=0\,.
\end{equation}

For fixed spin $J=\pm 1$, the Kerr-Schild vector corresponds to two of our flat null tetrad members
\begin{equation}
    \KSvec_{\mu;+}=m_{\mu}\,, \quad  \KSvec_{\mu;-}=\bm_{\mu}\,.
\end{equation}
To construct the null tetrad with respect to the Kerr-Schild metric $g^\Delta_{\mu\nu;J}$ for fixed helicity $J=\pm 2$, observe that the only inner product that changes is that of the Kerr-Schild vector of opposite helicity $-J=\mp 1$, namely $g^\Delta_{\mu\nu;J} \KSvec^\mu_{-J}\KSvec^\nu_{-J}= \varphi^\Delta$.
It is straightforward to complete the curved space null tetrad via a suitable linear combination of $\{\KSvec_+,\KSvec_-\}$ 
\begin{equation}
    \KSvec_{\mu;-J}^\Delta \equiv \KSvec_{\mu;-J}+\frac{\varphi^\Delta}{2} \KSvec_{\mu;J}\,.
\end{equation}
Notice, however, that raising the index on this mixed helicity vector with the inverse Kerr-Schild metric $g^{\Delta \, \mu \nu}_J$ (i.e. $   \KSvec_{-J}^{\Delta\,\mu} \equiv \KSvec^\mu_{-J}-\frac{\varphi^\Delta}{2} \KSvec^\mu_{J}$)
no longer yields the same result as raising the index with $\eta^{\mu\nu}$, as was the case for the Kerr-Schild vectors of definite helicity.
In terms of the tetrad
\begin{equation}\label{gTetrad}
    \{l,n,\KSvec_J,\KSvec^\Delta_{-J}\},
\end{equation}
the Kerr-Schild metric can be expressed as
\begin{equation}\label{KSmetricTetrad}
    g^\Delta_{\mu\nu;J}=-l_\mu n_\nu-n_\mu l_\nu+\KSvec_{\mu;J}\KSvec^\Delta_{\nu;-J}+\KSvec^\Delta_{\mu;-J}\KSvec_{\nu;J}\,.
\end{equation}

The Maxwell field strength for the gauge field in the Minkowski background is
\begin{equation}\label{KSFieldStrength}
F^\Delta_{\mu\nu;J}=(\Delta-1)\Big(l_\mu \KSvec_{\nu;J}-l_\nu \KSvec_{\mu;J}\Big) \varphi^\Delta
\,,
\end{equation}
and the Riemann tensor for the Kerr-Schild metric is
\begin{equation}\label{KSRiemann}
 R^\Delta_{\mu\nu \rho\sigma;J}
  =-\frac{1}{2} \Delta(\Delta-1)(l_\mu \KSvec_{\nu;J}-l_\nu \KSvec_{\mu;J}) (l_\rho \KSvec_{\sigma;J}-l_\sigma \KSvec_{\rho;J}) \varphi^\Delta
\,.
\end{equation}
The $\Delta$-dependent factors in~\eqref{KSFieldStrength} and~\eqref{KSRiemann} ensure the curvatures vanish when the conformal primaries $A^\Delta_{\mu;J}$ and $h^\Delta_{\mu\nu;J}$ reduce to pure gauge and diffeomorphism terms, respectively, when $\Delta=1$ and $\Delta=0,1$.
Furthermore, the Ricci tensor for the Kerr-Schild metric~\eqref{gKerrSchild} vanishes
\begin{equation}\label{RDelta}
 R^\Delta_{\mu\nu;J}=0\,.
\end{equation}
This can be seen using the properties of the null tetrad implied by $q^2=q\cdot \epsilon_J=0$; or, alternatively, by starting from the famous property that Kerr-Schild metrics linearize the Ricci tensor with mixed indices, as in~\eqref{KSRicci}, and using the null and geodesic properties of the Kerr-Schild vector~\eqref{nullgeodesick}.
Hence the Minkowski background ``perturbed'' by the spin-two conformal primary $h^{\Delta}_{J=\pm2}$, originally constructed as a linearized solution, is actually an exact solution to the vacuum Einstein equations! 

Because of~\eqref{RDelta}, the Riemann tensor~\eqref{KSRiemann} is equivalent to the Weyl tensor as expected for exact gravitational wave solutions in the vacuum. 
Out of the five Newman-Penrose Weyl scalars only one is non-vanishing\footnote{Note that these definite helicity solutions have complexified Newman-Penrose scalars, so $\bar{\Psi}_i$ and $\Psi_i$ are independent.}
\begin{equation}\label{NPPsi4}
    \Psi_4=C^\Delta_{\mu\nu\rho\sigma;J} n^\mu \KSvec^\nu_{-J}n^\rho \KSvec^\sigma_{-J}=-\frac{1}{2}\Delta(\Delta-1) \varphi^\Delta\,.
\end{equation}
Hence the conformal primary $h^{\Delta}_{J=\pm2}$ gives rise to a Petrov type {\bf N} solution to Einstein's equations.

\subsection{Conformal Shadow Primaries}\label{sec:cpwsh}

The shadow transformed conformal primary wavefunctions also satisfy a Kerr-Schild double copy
\begin{equation}
    \tA^{\Delta}_{\mu;J}=\KSvec_{\mu;J}\tvarphi^{\Delta}\,, \quad \th^{\Delta}_{\mu\nu;J}=\KSvec_{\mu;J}\KSvec_{\nu;J}\tvarphi^{\Delta}\,,
\end{equation}
for the same Kerr-Schild vector~\eqref{KSvector}, which is null and geodesic with respect to the Kerr-Schild shadow metric
\begin{equation}\label{SHKSmetricTetrad}
    \tg^{\Delta}_{\mu\nu;J}=\eta_{\mu\nu}+\KSvec_{\mu;J}\KSvec_{\nu;J}\tvarphi^{\Delta}\,.
\end{equation}
The expressions for the field strength and Riemann tensor, which are rather complicated, can be vastly simplified by adopting a modified version of the null tetrad~\eqref{gTetrad}, namely
\begin{equation}
    \label{SHgTetrad}
    \{l,n,\KSvec_J,\tKSvec^{\Delta}_{-J}\}
\end{equation}
where
\begin{equation}
    \tKSvec_{-J}^{\Delta\,\mu} \equiv \KSvec^\mu_{-J}-\frac{\tvarphi^{\Delta}}{2} \KSvec^\mu_{J}\,.
\end{equation}
In terms of this tetrad, the field strength for the Maxwell shadow gauge field is
\begin{equation}\label{SHKSFieldStrength}
\tF^{\Delta}_{\mu\nu;J}=(\Delta-1)\Big(n_\mu \KSvec_{\nu;J}-n_\nu \KSvec_{\mu;J}\Big) \frac{2}{X^2}\tvarphi^{\Delta}
\,,
\end{equation}
and the Riemann tensor for the Kerr-Schild shadow metric~\eqref{SHKSmetricTetrad} is
\begin{equation}\label{SHKSRiemann}
 \tR^{\Delta}_{\mu\nu \rho\sigma;J}
  =-\frac{1}{2} (\Delta-1)(\Delta-2)(n_\mu \KSvec_{\nu;J}-n_\nu \KSvec_{\mu;J}) (n_\rho \KSvec_{\sigma;J}-n_\sigma \KSvec_{\rho;J}) \left(\frac{2}{X^2}\right)^2\tvarphi^{\Delta}
\,.
\end{equation}
Note that the curvatures have the same tensor structure predicted by the map~\eqref{notsh}.  Meanwhile, the numerical prefactor matches the behavior expected from table~\ref{table:shadowGoldstone}.
The conformal shadow primaries $\tA^{\Delta}_{\mu;J}$ and $\th^{\Delta}_{\mu\nu;J}$ become pure gauge when, respectively, $\Delta=1$ and $\Delta=1,2$.
The Ricci tensor vanishes
\begin{equation}
    \tR^{\Delta}_{\mu\nu;J}=0\,,
\end{equation}
and the sole non-vanishing Weyl scalar is
\begin{equation}\label{SHNPPsi4}
    \Psi_4=\tC^{\Delta}_{\mu\nu\rho\sigma;J} l^\mu \KSvec^\nu_{-J}l^\rho \KSvec^\sigma_{-J}=-\frac{1}{2}(\Delta-1)(\Delta-2) \left(\frac{2}{X^2}\right)^2 \tvarphi^{\Delta}\,.
\end{equation}
The conformal shadow primary $\th^{\Delta,\pm}_{J=\pm2}$ is thus also an exact solution to Einstein's equations of Petrov type {\bf N}. 

Notice that the roles of principle null directions $l$ and $n$ in~\eqref{SHNPPsi4} are exchanged as compared to~\eqref{NPPsi4}. Hence, there is a sense in which the primaries $h^{\Delta}_{J=\pm2}$ and $\th^{\Delta}_{J=\pm 2}$ for fixed $J$ and reference direction are counter-propagating.

\subsection{Conformally Soft Double Copy}\label{sec:cscpw}

\subsubsection{Goldstone Modes}
 The field strengths and Riemann tensors for the primaries,~\eqref{KSFieldStrength} and~\eqref{KSRiemann}, and for their shadows,~\eqref{SHKSFieldStrength} and~\eqref{SHKSRiemann}, vanish for $\Delta=1$. At this value of the conformal dimension, the conformal primaries and their shadows degenerate and become the Goldstone modes of spontaneously broken large gauge symmetry
\begin{equation}\label{A1}
    \tA^1_{\mu;J}=A^1_{\mu;J}=\nabla_\mu \alpha^1_J\,,\quad \text{with} \quad \alpha^{1}_J= -\frac{\epsilon_{J}\cdot X}{q\cdot X}\,,
\end{equation}
and of spontaneously broken diffeomorphism symmetry
\begin{equation}
    \th^1_{\mu\nu;J}=h^1_{\mu\nu;J}=\nabla_{(\mu} \xi^1_{\nu);J}\,, \quad \text{with} \quad \xi^{1}_{\mu;J}=-\frac{1}{2}  \frac{\epsilon_{J} \cdot X}{q\cdot X}\left(\epsilon_{\mu;J} -\frac{1}{2} q_\mu\frac{ \epsilon_{J}\cdot X}{q\cdot X}\right)\,.
\end{equation}
From the Kerr-Schild double copy 
\begin{equation}
    h^1_{\mu\nu;J}=\KSvec_{\mu;J} A^1_{\nu;J}=\KSvec_{\mu;J}\KSvec_{\nu;J}\varphi^1\,,
\end{equation}
it follows that the generator of BMS symmetry $\xi^1_\mu$ is related to the generator of large $U(1)$ Kac-Moody symmetry $\alpha^1$.
 In this sense, BMS symmetry is a double copy of large gauge symmetry~\cite{Huang:2019cja,Alawadhi:2019urr}. Since all Weyl scalars vanish, the Kerr Schild metrics~\eqref{gKerrSchild} and~\eqref{SHKSmetricTetrad} become Petrov type {\bf O}, as expected for a pure diffeomorphism.

\subsubsection{Conformally Soft Primaries}

A linear combination of conformal primaries and their shadows has been constructed in~\cite{Donnay:2018neh} in the {\it conformally soft} limit $\Delta=1$. The conformally soft photon is given by
\begin{equation}\label{ACS}
  A^{\rm CS}_{\mu;J} = \Big[\Theta(X^2)+\log[X^2](q \cdot X) \delta(q\cdot X) \Big] A^{\rm 1}_{\mu;J}\,,
 \end{equation}
 whose field strength on the flat background is
 \begin{equation}
   F^{\rm CS}_{\mu\nu;J}=2\left(X_\mu A^1_{\nu;a}-X_\nu A^1_{\mu;J}\right) \Big[\delta(X^2)+\frac{q\cdot X}{X^2} \delta(q\cdot X)\Big]\,.
 \end{equation}
 The conformally soft graviton is
\begin{equation}\label{hCS}
 h^{\rm CS}_{\mu\nu;J}=\left[\Theta\left(X^2\right)+\log[X^2](q \cdot X) \delta(q\cdot X)\right] h^{\rm 1}_{\mu\nu;J}\,,
\end{equation}
and the Riemann tensor for the perturbed metric 
\begin{equation}\label{gCSKerrSchild}
 g^{\rm CS}_{\mu\nu;J}=\eta_{\mu\nu}+h^{\rm CS}_{\mu\nu;J}\,,
\end{equation}
is given by
\begin{equation}
 \badat{4}\label{CSreimann}
  R^{\rm CS}_{\mu\nu\rho\sigma;J}&=-\Big\{\Big[\Big(\eta_{\rho\mu}-2\frac{X_\rho X_\mu}{X^2}\Big) h^1_{\sigma\nu;J} - (\rho \leftrightarrow \sigma)\Big]-(\mu \leftrightarrow \nu)\Big\}\Big[\delta(X^2)+\frac{q\cdot X}{X^2} \delta(q\cdot X)\Big]\,.
 \eadat
\end{equation}
The Ricci tensor vanishes
\begin{equation}
 R^{\rm CS}_{\mu\nu;J}=0\,,
\end{equation}
and the conformally soft graviton thus gives rise to an exact solution to the vacuum Einstein equations as well. 

Indeed, the conformally soft gravitons and photons also satisfy a Kerr-Schild double copy 
\begin{equation}
 h^{\rm CS}_{\mu\nu;J}=\KSvec_{\mu;J} A^{\rm CS}_{\nu;J}=\KSvec_{\mu;J}\KSvec_{\nu;J} \varphi^{\rm CS}\,.
\end{equation}
with Kerr-Schild vector~\eqref{KSvector}, which is also null and geodesic with respect to~\eqref{gCSKerrSchild}.
This in turn defines the conformally soft scalar
\begin{equation}\label{CSscalar}
 \varphi^{\rm CS}=\left(\Theta\left(X^2\right)+\log[X^2](q \cdot X) \delta(q\cdot X)\right) \varphi^1\,,
\end{equation}
which solves the massless Klein-Gordon equation in the flat background and transforms as a $\Delta=1$ conformal primary wavefunction. 

In terms of the null tetrad $\{l,n,\KSvec_J,\KSvec^{\rm CS}_{-J}\}$ where 
\begin{equation}
    \KSvec^{\rm CS}_{-J} \equiv \KSvec^\mu_{-J}-\frac{\varphi^{\rm CS}}{2} \KSvec^\mu_{J}\,,
\end{equation}
the Riemann tensor can be expressed as
\begin{equation}
 \badat{4}
  R^{\rm CS}_{\mu\nu\rho\sigma;J}
   &=\Big[\frac{X^2}{2}(l_\rho \KSvec_{\sigma;J}-l_\sigma \KSvec_{\rho;J}) (l_\mu \KSvec_{\nu;J}-l_\nu \KSvec_{\mu;J})\\
   &\quad +\frac{2}{X^2}(n_\rho \KSvec_{\sigma;J}-n_\sigma \KSvec_{\rho;J}) (n_\mu \KSvec_{\nu;J}-n_\nu \KSvec_{\mu;J})\Big]\Big[\delta(X^2)+\frac{q\cdot X}{X^2} \delta(q\cdot X)\Big]\varphi^1\,.
 \eadat
\end{equation}
We thus obtain a superposition of solutions with the tensor structures we found for the primaries $h^{\Delta}_{J=\pm2}$ and $\th^{\Delta}_{J=\pm2}$, but with singular support at the loci $X^2=0$ and $q\cdot X=0$. 
Up to symmetries of the Weyl tensor, the only non-zero Newman-Penrose scalars are given by the contractions
\begin{equation}
 \badat{4}
  C^{\rm CS}_{\mu\nu\rho\sigma;J}n^{\mu}\KSvec_{-J}^{\rm CS\, \nu} n^{\rho}\KSvec_{-J}^{\rm CS\, \sigma} 
  &=\frac{X^2}{2}\Big[\delta(X^2)+\frac{q\cdot X}{X^2} \delta(q\cdot X)\Big]\varphi^1\,,\\
 \eadat
\end{equation}
and
\begin{equation}
 \badat{4}\label{eqp2}
  C^{\rm CS}_{\mu\nu\rho\sigma;J}l^{\mu}\KSvec_{-J}^{\rm CS\,\nu} l^{\rho}\KSvec_{-J}^{\rm CS\,\sigma}
  &=\frac{2}{X^2}\Big[\delta(X^2)+\frac{q\cdot X}{X^2} \delta(q\cdot X)\Big]\varphi^1\,.\\
 \eadat
\end{equation}
From this result, it appears that the CS modes are not of type {\bf N}.  However, there is a subtlety here.  
In the limit that $q\cdot X\rightarrow 0$, the tetrad components $l^\mu$ and $n^\mu$ diverge and, so long as $X^2\neq0$, are both approximately proportional to $q^\mu$.  

If we are careful about the space of test functions, we can elect to keep only the leading order terms at each locus of non trivial support.
\begin{equation}
 \badat{4}\label{eqrfin}
  R^{\rm CS}_{\mu\nu\rho\sigma;J}
   &\cong\Big[\frac{2}{X^2}(X_\rho \KSvec_{\sigma;J}-X_\sigma \KSvec_{\rho;J}) (X_\mu \KSvec_{\nu;J}-X_\nu \KSvec_{\mu;J})\delta(X^2)\\
   &~~+(q_\rho \KSvec_{\sigma;J}-q_\sigma \KSvec_{\rho;J}) (q_\mu \KSvec_{\nu;J}-q_\nu \KSvec_{\mu;J})\delta(q\cdot X)\Big]\varphi^1\,
 \eadat
\end{equation}
which, although singular as a distribution, is type {\bf N}.

\section{Generalized Conformal Primaries}\label{sec:gencpw}
The previous section showed that conformally soft modes lead to distributional curvatures.  This suggests we may also want to put more thought into allowing distributional versions of our conformal primary wavefunctions.
Another issue we have avoided up to this point is that removing the $i\varepsilon$ regulator opens up the possibility of sources. We will address both of these topics by introducing the notion of a {\it generalized conformal primary wavefunction}.

\subsection{Generalized Scalar}\label{sec:gencpws}
Since we are ultimately interested in constructing non-trivial backgrounds for celestial CFT, we consider only bosonic fields in this section.  As we saw from the double copy result, it is convenient to build up the spin-1 and spin-2 wavefunctions in terms of the scalar. 
\vspace{1em}

\noindent{\bf Definition:} A {\it generalized conformal primary scalar} is a wavefunction on $\mathbb{R}^{1,3}$ which transforms under $SL(2,\mathbb{C})$ as a conformal primary of dimension $\Delta$.
\be\label{gens}
\varphi^{gen}_{\Delta}\Big(\Lambda^\mu_\nu X^\nu;\frac{a w+b}{cw+d},\frac{{\bar a} \bw+{\bar b}}{{\bar c}\bw+{\bar d}}\Big)=|cw+d|^{2\Delta}\varphi^{gen}_{\Delta}(X^\mu;w,\bw).
\ee
That's it.  We don't impose any condition on the equation of motion.  

Wavefunctions which are analytic in $X^\mu$ take the form
\be\label{genphi}
\varphi^{gen}_{\Delta}=f(X^2)\frac{1}{(-q\cdot X)^\Delta}\,,
\ee
for some $f(X^2)$, while for distributional wavefunctions we should allow the $\varphi^\Delta$ factor to vary as well.  For instance
\be\label{dfn}
\lim\limits_{\varepsilon\rightarrow 0}i(\varphi^{1,+}-\varphi^{1,-})=2\pi\delta(q\cdot X)\,.
\ee
While singular, one may also start from the distribution
\be
\varphi^{gen}_{\Delta}=\frac{1}{(-q\cdot X)^{\Delta-1}}\delta(q\cdot X)\,,
\ee
and other distributional combinations, depending on the use case and the space of test functions one wants to allow.  We won't judge.

\begin{figure}[ht]
    \centering
    \includegraphics{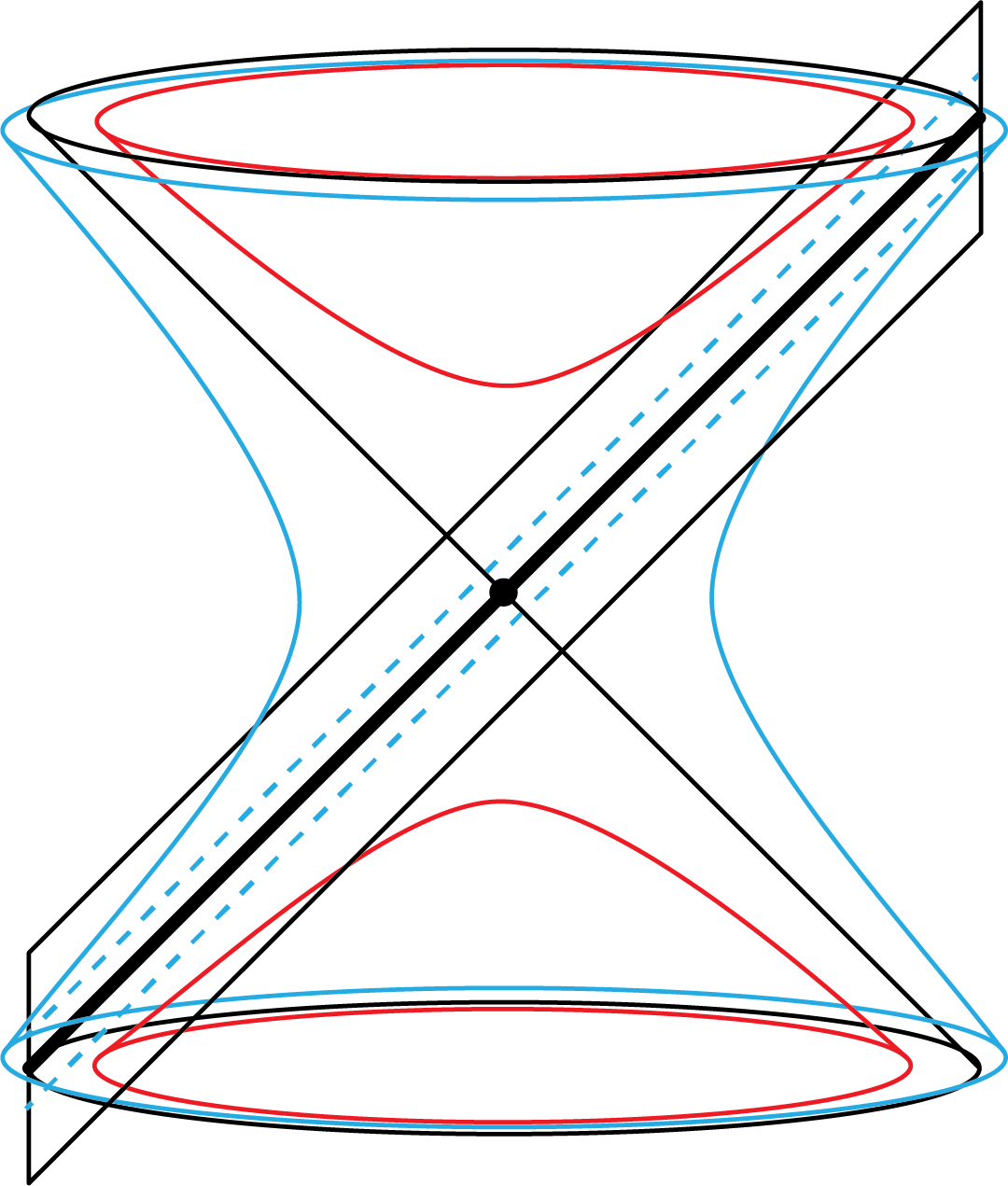}
    \caption{Loci of support consistent with conformal covariance. Wavefunctions obeying~\eqref{gens} are naturally written in terms of data on the hyperbolic foliation of Minkowski space, multiplied by a homogeneous function in the distance from the hyperplane $q\cdot X=0$.  Wavefunctions with restricted support will have that support on a union of some leaves of the hyperbolic foliation with constant $X^2$, on the hyperplane $q\cdot X=0$, or on an intersection thereof (bold and dashed curves).}
    \label{support}
\end{figure}

The $f(X^2)$ term can also be a distribution.  If support is restricted to one of the Milne regions, one can also multiply by $\Theta(X^0)$ and not ruin the property~\eqref{gens}.  The key point to remember is that we have the hyperbolic foliation~\cite{deBoer:2003vf} and the hyperplane $q\cdot X=0$ at our disposal.  The leaves of the foliation~\cite{deBoer:2003vf} are defined by $\delta(X^2-\tau^2)$, and $\delta(X^2+\tau^2)\Theta(\pm X^0)$ for real $\tau$. Figure~\ref{support} shows a schematic of the submanifolds on which a distributional generalized scalar wavefunction can have support.

To be more concrete, consider the case $w=0$ so that $q^\mu=(1,0,0,1)$ and let the 3D volume projected onto the page in figure~\ref{support} be the hypersurface $y=0$. The bold line in figure~\ref{support} is the wordline of a massless point particle, while the dashed lines are the loci $\{t=z,x^2+y^2=\tau^2>0\}$, i.e. we are seeing a 3D cross section of a worldsheet in 3+1D, rather than two worldlines.

\subsubsection{Source Interpretation}
While we do not impose any conditions on the equation of motion, we do stress that the quantity $\Box \varphi^{gen}_{\Delta}$ should retain the same conformal transformation properties as $\varphi^{gen}_{\Delta}$, so that if it is non-vanishing it also has conformal dimension $\Delta$ and spin $J=0$. We can use this feature to reach more general distributions than the examples given above. For instance starting from~\eqref{dfn}, wavefunctions of the form $\Box^n\delta(q\cdot X)$ will also have generalized conformal dimension 1.  Alternatively, we can look at solutions where the support of the source is of higher codimension than the support of the wavefunction.  The locus of support for the source should also possess the structure described in figure~\ref{support}, and should be a subset of the support of the wavefunction itself.

\subsubsection{Off-shell Interpretation}
One can also use this construction to look for off-shell solutions, e.g to the Klein-Gordon equation
\be\label{kgpm}
\Box \varphi^{gen}_\Delta=\pm m^2\varphi^{gen}_\Delta\,.
\ee
In~\cite{Pasterski:2016qvg}, we were interested in positive and negative frequency solutions to the standard real-mass (i.e. $+$~sign in~\eqref{kgpm}) Klein-Gordon equation
\be \label{asf}
\phi^\pm_{\Delta,m}( X^\mu;w,\bar w)  = \frac{4\pi}{im} \frac{(\sqrt{-X_\pm^2} )^{\Delta-1}}{(-q\cdot X_\pm)^\Delta} K_{\Delta-1} \left( im\sqrt{-X_\pm^2} \right) \,,
\ee
where we picked the Bessel function with good falloff behavior and restored an $i\varepsilon$ prescription to avoid singularities/sources on the submanifold of real $X^\mu$. To arrive at~\eqref{asf} we actually, somewhat ironically, solved~\eqref{kgpm} for the ``wrong sign'', and then analytically continued away from \mbox{$m\in -i \mathbb{R}_+$}. Here, we are essentially pointing out that we can use a hyperbolic foliation of off-shell momentum space to capture off-shell exchanges if we take this analytic continuation literally.  Dealing with strictly massless particles, we would expect cutting rules to return us to the on-shell Mellin (or shadow Mellin) solutions.  While not specifically looking for off-shell solutions, the machinery one would want to use is likely related to the auxiliary exchange modes studied in~\cite{Law:2020xcf}.

\subsubsection{Generalized Conformally Soft Scalar}

In looking for examples of generalized conformal scalars, the case $\Delta=1$ relevant in section~\ref{sec:cscpw} is subtle and requires special care. We will start by looking at solutions that satisfy the massless Klein-Gordon equation away from singular loci, and then see how the $\Delta\rightarrow 1$ limit can lead to solutions other than one considered in~\cite{Donnay:2018neh}, if one avoids imposing the standard $i\varepsilon$ prescriptions.  

Starting with the factorized ansatz 
\be\label{gen}
\varphi^{gen}_\Delta=f(X^2)S^\Delta(q\cdot X)\,,
\ee
where 
\begin{equation}
    X^\mu\p_\mu S^\Delta=-\Delta S^\Delta\,,
\end{equation}
one can show that~\cite{Pasterski:2017kqt}
\begin{equation}\label{eqsol1}
   \Box \varphi^{gen}_\Delta=0~~\Leftrightarrow ~~X^2 f''(X^2)-(\Delta-2) f'(X^2)=0\,.
\end{equation}
For generic $\Delta$, this differential equation has the solution
\begin{equation}\label{eqsol2}
    f=c_1+c_2 (X^2)^{\Delta-1}\,,
\end{equation}
while for $\Delta=1$, it has the solution
\begin{equation}\label{log}
    f=c'_1+c'_2 \log(X^2)\,.
\end{equation}
One can also obtain this log solution from a limit of the power law ones (which was implicitly done in~\cite{Donnay:2018neh}).  It is straightforward to see that the multiplicative function
\be
\log(X^2)=\lim\limits_{\Delta\rightarrow1}\frac{(X^2)^{\Delta-1}-1}{\Delta-1}
\ee
arises via the limit
\begin{equation}\label{logmode}
   \varphi^{\log;\pm} \equiv \lim\limits_{\Delta\rightarrow1}\p_\Delta (\varphi^{\Delta,\pm}+\tvarphi^{2-\Delta,\pm})
   =-\frac{\log(X_\pm^2)}{(-q\cdot X_\pm)}\,.
\end{equation}
The conformally soft mode of~\cite{Donnay:2018neh}, which appears above as the Kerr-Schild zeroth copy~\eqref{CSscalar}, is constructed via
\be\label{varphiCS}
\varphi^{CS}=\frac{1}{2\pi i}(\varphi^{\log,+}-\varphi^{\log,-})=\left(\Theta\left(X^2\right)+\log[X^2](q \cdot X) \delta(q\cdot X)\right) \varphi^1.
\ee

Looking back at the definition~\eqref{gens}, it becomes clear that either one of these terms individually meets our requirement for a generalized conformal scalar of conformal dimension $\Delta=1$.  We can re-address the concerns we raised at the end of section~\ref{sec:cscpw} about the singular behavior and the Petrov type of the solution by restricting each of the terms in~\eqref{varphiCS} separately. Defining
\be\label{pcsp}
\varphi^{CS'}\equiv \Theta(X^2)\frac{1}{-q\cdot X}\,, \quad \varphi^{CS''}\equiv\log[X^2]\delta(q\cdot X)\,,
\ee
one can check that these wavefunctions are separately responsible for the two terms in~\eqref{eqrfin}.
At the end of this section, we will consider an alternative Kerr-Schild vector for the double copy construction that uses the wavefunction $\varphi^{CS''}$  to construct the Aichelburg-Sexl ultra-boost.

\subsection{Generalized Vector}\label{sec:gencpwv}
Building on the pattern we started for the generalized conformal scalar wavefunction, we will now consider the analogue for integer spin.  We start with minimal assumptions, only enforcing conformal covariance, and then examine, in turn, how these solutions are restricted by the tracelessness, the equations of motion, and the gauge choices we had imposed to select the corresponding conformal primary wavefunctions.  We begin with the vector field case.
\vspace{1em}

\noindent{\bf Definition:} A {\it generalized conformal primary vector} is a wavefunction on $\mathbb{R}^{1,3}$ which transforms under $SL(2,\mathbb{C})$ as a conformal primary of dimension $\Delta$ and spin $J$
\be\label{gena}
A^{gen}_{\Delta,J;\mu}\Big(\Lambda^\mu_\nu X^\nu;\frac{a w+b}{cw+d},\frac{{\bar a} \bw+{\bar b}}{{\bar c}\bw+{\bar d}}\Big)=(cw+d)^{\Delta+J}({\bar c}\bw+{\bar d})^{\Delta-J} \Lambda_\mu^{~\nu} A^{gen}_{\Delta,J;\nu}(X^\mu;w,\bw)\,.
\ee
While for the spin-1 radiative solutions we had $|J|=1$, we will see that generically rank-$s$ tensor fields in 3+1D can have 2D conformal spins $|J|\le s$, matching what one would expect for decomposing the spin states of a massive particle -- after all we are not imposing the massless equations of motion at this stage. One can view the conformal spin as measuring the $3+1$D spin along the axis parallel to the reference direction.  See, for instance, the examination of the massless limit of massive spin-1 primaries in~\cite{ss}, or the general massive classification~\cite{Law:2020tsg}. 

Using~\eqref{tetrad} and their $SL(2,\mathbb{C})$ transformation~\eqref{tetradcov}, we see that our desired $J=0$ solutions take the form
\be\label{agen0}
A^{gen}_{\Delta,0;\mu}= l_\mu \varphi_{\Delta}^{gen,1}+ n_\mu \varphi_{\Delta}^{gen,2}\,,
\ee
where the two generalized conformal primary scalar wavefunctions that appear are generally different, and any one of them can vanish.  Meanwhile for $J=\pm1$ we have
\be\label{agen1}
A^{gen}_{\Delta,+1;\mu}=m_\mu \varphi_{\Delta}^{gen},~~A^{gen}_{\Delta,-1;\mu}={\bar m}_\mu \varphi_{\Delta}^{gen}\,,
\ee
again for arbitrary generalized conformal primary scalar wavefunctions.  Because the inner products of the tetrad elements~\eqref{eq:iptetrad} are just $\pm 1$ or zero, this is all the freedom we get. 

We note the radial gauge condition imposes one constraint on~\eqref{agen0}, but is automatically satisfied by~\eqref{agen1}.  Spatial derivatives hitting $\varphi^{gen}_\Delta$ will be in the span of $\{X^\mu, q^\mu\}$. Furthermore
\be
X\cdot l=-1\,,~~~X\cdot n=\frac{X^2}{2}\,,~~~X\cdot m=0\,,~~~X\cdot \bm=0\,,
\ee
while derivatives of the tetrad take the form
\be
\p_\mu l_\nu= l_\mu l_\nu\,,~~\p_\mu n_\nu=\eta_{\mu\nu}+ n_\mu l_\nu\,,~~\p_\mu m_\nu= m_\mu l_\nu\,,~~\p_\mu \bm_\nu= \bm_\mu l_\nu\,,
\ee
from which one can check
\be
\Box l^\mu=0\,,~~~\Box n^\mu=2l^\mu\,,~~~\Box m^\mu=0\,,~~~\Box \bar{m}^\mu=0\,.
\ee
This allows us to compute the constraints on the generalized conformal scalar wavefunctions that appear when imposing the harmonic gauge condition or spin-1 equations of motion.  We will only do so explicitly for the analytic generalized conformal primary scalar~\eqref{genphi}, where the only freedom is the choice of $f(X^2)$.   Table~\ref{Aconst} summarizes the results.  When harmonic gauge is not imposed, the entries of this table need to be combined as in~\eqref{lineom} to identify the current.
\begin{table}[ht]
\renewcommand*{\arraystretch}{1.3}
\centering
    \begin{tabular}{l|l|l|l}
         & $X^\mu A_\mu$ &  $\p^\mu A_\mu$ & $\Box A_\mu$ \\  
         \hline
        $A^{gen}_{\Delta,+1}$& 0 & 0 &$4[(2-\Delta)f'+X^2f'']m_\mu\varphi^\Delta$\\
      $A^{gen}_{\Delta,0}$& $[-f_1+\frac{X^2}{2}f_2]\varphi^\Delta$ & $[-2f'_1+(3-\Delta)f_2+X^2f'_2]\varphi^\Delta$&$\Big\{4[(1-\Delta)f_1'+X^2f_1''+\frac{1}{2}f_2]l_\mu$\\
        &&&~~~$+4[(3-\Delta)f_2'+X^2f_2'']n_\mu\Big\}\varphi^\Delta$\\
    \end{tabular}
    \caption{Modes relevant for various constraints one can impose on a generalized conformal primary vector field. The allowed forms for negative $J$ solutions are related by complex conjugation.}
    \label{Aconst}
\end{table}

For the $J=\pm1$ examples, we see what would be the non-shadow and shadow modes arising from the second order differential equation for $f$ in the last column.  For the $J=0$ case, we have two functions constrained by a coupled system of second order equations. Trying to impose the same gauge fixing we used for the radiative modes makes the system highly constrained. For instance, table~\ref{Aconst} shows we would need $f_1=\frac{X^2}{2}f_2$ to obey the radial gauge condition. The harmonic gauge condition then forces $\Delta=2$.  The last column then tells us that we have a harmonic solution for \be
f_2'+X^2 f_2''=0~\Rightarrow f_2=c_1+c_2\log[X^2]
\ee
so that there are two independent $\Delta=2$, $J=0$ harmonic and radial gauge generalized conformal primary vectors given by 
\be
A_\mu=[n_\mu+\frac{X^2}{2}l_\mu](c_1+c_2\log[X^2])\varphi^2
\ee
which satisfy the free Maxwell equations almost everywhere.

\subsection{Generalized Tensor}\label{sec:gencpwt}
We can now see the pattern one would need to build up an arbitrary symmetric rank-$s$ tensor:  one constructs superpositions of symmetric products of tetrads with definite conformal spin $|J|\le s$  multiplied by a generalized conformal primary scalar of weight $\Delta$ (allowing an independent solution for each independently allowed tensor factor). We will now demonstrate the rank-2 example relevant to generalizing our metric solutions.
\vspace{1em}

\noindent{\bf Definition:} A {\it generalized conformal primary metric} is a wavefunction on $\mathbb{R}^{1,3}$ which transforms under $SL(2,\mathbb{C})$ as a conformal primary of dimension $\Delta$ and spin $J$
\be\label{genh}
h^{gen}_{\Delta,J;\mu\nu}\Big(\Lambda^\mu_\nu X^\nu;\frac{a w+b}{cw+d},\frac{{\bar a} \bw+{\bar b}}{{\bar c}\bw+{\bar d}}\Big)=(cw+d)^{\Delta+J}({\bar c}\bw+{\bar d})^{\Delta-J} \Lambda_\mu^{~\sigma}\Lambda_\nu^{~\rho} h^{gen}_{\Delta,J;\sigma\rho}(X^\mu;w,\bw)
\ee
and is symmetric under exchange of the $3+1$D indices.

Using~\eqref{tetrad} and their $SL(2,\mathbb{C})$ transformation properties~\eqref{tetradcov}, we see that our desired $J=0$ solution takes the form
\be\label{hgen0}
h^{gen}_{\Delta,0;\mu\nu}= l_\mu l_\nu \varphi_{\Delta}^{gen,1}+ n_\mu n_\nu \varphi_{\Delta}^{gen,2}+(l_\mu n_\nu+n_\mu l_\nu) \varphi_{\Delta}^{gen,3}+\eta_{\mu\nu}\varphi_{\Delta}^{gen,4}\,,
\ee
where the four generalized conformal primary scalar wavefunctions that appear are generally different, and any one of them can vanish.  Meanwhile for $J=+1$ we have
\be\label{hgen1}
h^{gen}_{\Delta,+1;\mu\nu}=(l_\mu m_\nu+m_\mu l_\nu) \varphi_{\Delta}^{gen,1}+(n_\mu m_\nu+m_\mu n_\nu) \varphi_{\Delta}^{gen,2}\,,
\ee
and the same for $J=-1$ with $m\mapsto \bar{m}$. Finally, for $J=+2$ we have
\be\label{hgen2}
h^{gen}_{\Delta,+2;\mu\nu}=m_\mu m_\nu \varphi_{\Delta}^{gen}\,,
\ee
and the same for $J=-2$ with $m\mapsto \bar{m}$.

We will now compute the constraints on the generalized conformal scalar wavefunctions that arise from imposing standard gauge constraints or equations of motion on the metric. Again, we will only do so explicitly for the analytic case~\eqref{genphi}, where the only freedom is the choice of $f(X^2)$.  The results are summarized in table~\ref{hconst}.  
\begin{table}[ht]
\renewcommand*{\arraystretch}{1.3}
    \centering
    \scalebox{0.9}{
    \begin{tabular}{l|l|l|l}
         & $\eta^{\mu\nu}h_{\mu\nu}$& $X^\mu h_{\mu\nu}$ &  $\p^\mu h_{\mu\nu}$ \\
         \hline
          $h^{gen}_{\Delta,+2}$&0& 0 & 0 \\
        $h^{gen}_{\Delta,+1}$&0&$[-f_1+\frac{X^2}{2}f_2]m_\nu\varphi^\Delta$  &  $[-2f'_1+(4-\Delta)f_2+X^2 f'_2] m_{\nu}\varphi^\Delta$\\
        $h^{gen}_{\Delta,0}$&$[-2f_3+4 f_4]\varphi^\Delta$&$\Big\{[-f_1+\frac{X^2}{2}(f_3-f_4)] l_\nu$&$\Big\{[-2f'_1+(2-\Delta)f_3+X^2 (f'_3-f'_4)+\Delta f_4] l_\nu $\\
        &&~~$+[\frac{X^2}{2}f_2-(f_3-f_4)] n_\nu\Big\}\varphi^\Delta$&$~~+[(4-\Delta)f_2+X^2 f'_2-2(f'_3-f'_4)]n_\mu\Big\}\varphi^\Delta $\\
    \end{tabular}
    }
    \caption{Modes relevant for various constraints one can impose on a generalized conformal primary metric.  The allowed forms for negative $J$ solutions are related by complex conjugation.}
    \label{hconst}
\end{table}
The fact that there are several allowed tensor structures makes the table entries rather long, so we have separated out evaluating the the d'Alembertian:
\begin{equation}
\label{hgenbox}
\scalemath{0.85}{
  \badat{4}
\Box h^{gen}_{\Delta,+2;\mu\nu}&=4[(2-\Delta)f'+X^2 f'']m_\mu m_\nu \varphi^{\Delta}\,,\\
\Box h^{gen}_{\Delta,+1;\mu\nu}&=4[(1-\Delta)f'_1+X^2 f''_1+f_2](l_\mu m_\nu+m_\mu l_\nu) \varphi^{\Delta}+4[(3-\Delta)f'_2+X^2 f''_2](n_\mu m_\nu+m_\mu n_\nu) \varphi^{\Delta}\,,\\
\Box h^{gen}_{\Delta,+0;\mu\nu}&=4[-\Delta f'_1+X^2 f''_1+f_3]l_\mu l_\nu \varphi^{\Delta}+4[(4-\Delta) f'_2+X^2 f''_2]n_\mu n_\nu \varphi^{\Delta}\\
&~~~+4[f_2+(2-\Delta) f'_3+X^2 f''_3](l_\mu n_\nu+n_\mu l_\nu) \varphi^{\Delta}+4[\frac{1}{2}f_2+(2-\Delta) f'_4+X^2 f''_4]\eta_{\mu\nu} \varphi^{\Delta}\,.
\eadat
}
\end{equation}
When we do not impose harmonic gauge or tracelessness, this result needs to be combined with the entries of table~\ref{hconst} into the linearized Einstein tensor to identify the matter stress tensor.
 
Recall, the point of this formalism was to set up a way to look at sourced solutions or non-radiative backgrounds on which to do scattering.   We have now reached the point where we have built enough machinery to dive into interesting particular examples.

\subsection{Boosted Black Hole, Shockwave, and Shift States}\label{sec:bhstates}

We will now show that combining our generalized conformal primary scalar with the Kerr-Schild double copy formalism allows us to identify the Aichelburg-Sexl ultra-boost metric with a generalized conformal primary metric of section~\ref{sec:gencpwt}, as well as several other examples.

We start with the generalized scalar mode $\varphi^{CS''}$, defined in~\eqref{pcsp} and appearing as one of the two terms in the conformally soft modes examined in~\cite{Donnay:2018neh}.  Because this mode appeared as a limit of source-free radiative modes, we expect its source to have support only where those modes had singularities in the $i\varepsilon\rightarrow0$ limit.  It is easier to identify this source if we pick a reference direction $q$.  Without loss of generality, we take $w=0$ so that $q^\mu=(1,0,0,1)$.  We then have
\be
\log[X^2]\delta(q_0\cdot X)=\log(x^2+y^2)\delta(t-z)\,,
\ee
then
\be\label{ptboost}
\badat{2}
\Box \left(\log[X^2]\delta(q_0\cdot X)\right)&=[(\p_x^2+\p_y^2)\log(x^2+y^2)]\delta(t-z)+\log(x^2+y^2)(-\p_t^2+\p_z^2)\delta(t-z)\\
&=2\pi\delta(x)\delta(y)\delta(t-z)\,,
\eadat
\ee
where we have used $(\p_t+\p_z)\delta(t-z)=0$ to kill the second term and employed the 2D Euclidean Green's function identity for the first term.  Promoting this to generic $q$, we get
\be
\Box \left(\log[X^2]\delta(q\cdot X)\right)=2\pi\int d\alpha \delta^{(4)}(X-\alpha q)\,.
\ee
Thus, the Klein-Gordon equation for this scalar mode is sourced by a massless point particle.

We will now use this scalar as a starting point for the double copy formalism.
In section~\ref{sec:CDCKS}, we observed that the conformal primary wavefunctions take the Kerr-Schild form where the polarization vectors $m$ or ${\bar m}$ serve as the Kerr-Schild vector.  Given what we saw in the previous subsections, we can try to generalize these solutions further by looking at other entries in our null tetrad to see if the Kerr-Schild form holds.

It is straightforward to check that the null geodesic requirement is satisfied by $l$ since
\be
l^\mu \nabla_\mu l_\nu=l^\mu \p_\mu l_\nu+l^\mu \Gamma^{\rho}_{\mu\nu} l_\rho=0\,.
\ee
The same also holds true for $q_\mu$.  Because the profiles which interest us are supported at $q\cdot X=0$, where the prefactors used to define our normalized tetrad diverge, we will use the $q_\mu$ in place of $l_\mu$ within the double copy formalism.  This does not get in the way of the classification we performed in the last section, but highlights the care with which one should approach excluding certain distributions.  The only thing to note here is that, while a double copy with $l_\mu$ relates fields of the same conformal dimension, a double copy with $q_\mu$ shifts the dimension down by one as the 3+1D spin goes up by one (but 2D spin stays constant).  Starting with the generalized conformally soft scalar $\varphi^{CS''}$, we land on a metric with conformal dimension $\Delta=-1$ and spin $J=0$.

\subsubsection{Aichelburg-Sexl}
The generalized conformal primary metric~\eqref{hgen0} of weight $\Delta=-1$ spin $J=0$
\be
l_\mu l_\nu \log[X^2]\delta(q\cdot X)(-q\cdot X)^2=q_\mu q_\nu\log[X^2]\delta(q\cdot X)
\ee
corresponds to the Aichelburg-Sexl ultra-boost metric~\cite{Aichelburg:1970dh}
\be\label{eq:as}
g_{\mu\nu}=\eta_{\mu\nu}-4G_N \alpha q_\mu q_\nu \log[X^2]\delta(q\cdot X)
\ee
with energy $E=\alpha q^0$.
The matter source is the bold line in figure~\ref{support}. In interesting recent work,~\cite{Cristofoli:2020hnk} derived this form of the metric starting from an off-shell amplitudes construction. We can proceed to match the other metrics considered in that paper to generalized conformal primaries.

\subsubsection{Ultra-boosted Schwarzschild-Tangherlini}
Reference~\cite{Cristofoli:2020hnk} also discussed the generalization of Aichelburg-Sexl to higher dimensions: the ultra-boosted Schwarzschild-Tangherlini metric studied in~\cite{Ferrari:1988cc}.
The comparison to results in~\cite{Cristofoli:2020hnk} does not stop there.  First, we point out that the analog of~\eqref{eqsol1} for $D>4$ ($d>2$ in~\cite{Pasterski:2017kqt}) implies the higher dimensional analog of ultra-boosted Schwarzschild corresponds to the shadow modes for a generalized conformally soft mode with $\Delta=1$
\be
(X^2)^{\Delta-\frac{d}{2}}\left(\frac{1}{(-q\cdot X_+)^{\Delta}}-\frac{1}{(-q\cdot X_-)^{\Delta}}\right)\Big |_{\Delta=1}=\frac{1}{(X^2)^{\frac{D-4}{2}}}\delta(q\cdot X)\,,
\ee
where $D=d+2$ in the notation of~\cite{Pasterski:2017kqt}.  We then arrive at
\be\label{eq:as22}
g_{\mu\nu}=\eta_{\mu\nu}+8\pi^{\frac{4-D}{2}}G_N \alpha q_\mu q_\nu\frac{\Gamma(\frac{D-2}{2})}{(D-4)(X^2)^{\frac{D-4}{2}}}\delta(q\cdot X),
\ee
again a $\Delta=-1$ spin $J=0$ conformal primary, now in a celestial CFT$_{D-2}$.

\subsubsection{Dray-'t~Hooft Planar Shell}
Consider the factorized form~\eqref{genphi}.  For the analytic conformal primary wavefunction~$\varphi^\Delta$, the class of generalized conformal primary wavefunctions is captured by the free function~$f(X^2)$.  In the last few subsections, however, we saw physically interesting solutions with distributional support on the locus $q\cdot X=0$.  For any such $S^\Delta(q\cdot X)$ as in~\eqref{gen}, one has a choice of $f(X^2)$.  In this and the following subsections, we will look at this type of generalization of the examples we have just considered.  Namely, we take 
\be\label{relf}
\varphi^{gen}_{\Delta=1}=f(X^2)S^{\Delta=1}(q\cdot X),~~~S^{\Delta=1}=\delta(q\cdot X)
\ee
and show that different $f(X^2)$ take us from the Aichelburg-Sexl metric to Dray-'t~Hooft's planar shell, to the ultra-boosted Kerr solution.  Moreover, this freedom to choose $f(X^2)$ is isomorphic to the freedom examined by Ferrari, Pendenza and Veneziano~\cite{Ferrari:1988cc}.  We will start with the specific example of Dray -'t~Hooft, demonstrate the generalization of~\cite{Ferrari:1988cc} corresponds to our classification here, then return to the specific example of the Kerr ultra-boost, which ties in directly to comments on spin memory and superrotation vacuum transitions with which we would like to close.

As pointed out in~\cite{Ferrari:1988cc}, for the case $f(X^2)=X^2$, we have
\be
\Box X^2\delta(q_0\cdot X)
=8\pi\delta(t-z)\,,
\ee
so that in place of the localized source~\eqref{ptboost} (bold line in figure~\ref{support}), there is a constant density of matter across the plane $q\cdot X=0$ (plane tangent to light cone in figure~\ref{support}).  This holds for arbitrary $D$ with the change $8\pi\mapsto 4(D-2)\pi$ coming from the the derivatives hitting $\sum_{i=1}^{d-1}x_i^2$. We see that the metric 
\be\label{eq:as3}
g_{\mu\nu}=\eta_{\mu\nu}-4\pi G_N \alpha q_\mu q_\nu\frac{X^2}{(D-2)}\delta(q\cdot X)
\ee
reproduces the Dray-'t Hooft plane wave solution~\cite{Dray:1985ie} with energy density $\rho=\alpha q^0$.

\subsubsection{Beam-like Gravitational Waves}
Ferrari, Pendenza, and Veneziano showed in~\cite{Ferrari:1988cc} that by changing the profile function $f(X^2)$, parameterizing the relative freedom~\eqref{relf}, one can build up an arbitrary beam.  In our language, functions that respect the hyperbolic foliation of~\cite{deBoer:2003vf} keep the correct conformal covariance properties.  Because the authors in~\cite{Ferrari:1988cc} are already restricted to the plane $q\cdot X=0$, they have $X^2=r^2$ where $r$ is the radial coordinate of a beam surrounding the null ray $X^\mu \propto q^\mu$.  The two dashed lines in figure~\ref{support} represent the intersection of the null hyperplane $q\cdot X=0$ and a hyperboloid of constant $X^2$, which only happens for $X^2>0$.  As explained in the discussion surrounding figure~\ref{support}, this surface is actually a worldsheet in the full $3+1$D spacetime of which the figure shows a cross section.

Sticking to the case $D=4$ here,~\cite{Ferrari:1988cc} points out that the metric
\be\label{eq:as2}
g_{\mu\nu}=\eta_{\mu\nu}-4 G_N \alpha f(X^2) q_\mu q_\nu \delta(q\cdot X)
\ee
describes the metric surrounding a beam of null matter with energy profile
\be
E(r)=\alpha q^0\sqrt{X^2}f'(X^2)\,.
\ee
The authors go on to examine geodesics in this background.  From the framework we have set up in this paper, we identify this metric as a generalized conformal primary metric of weight $\Delta=-1$, spin $J=0$, and double copy form.  Moreover, the space of solutions~\cite{Ferrari:1988cc} considers is precisely the degree of freedom we have in picking a generalized conformal primary scalar of dimension $\Delta=1$ for the choice of $S^{\Delta=1}$ in~\eqref{relf}. This encompasses the limiting Aichelburg-Sexl/Schwarzschild-Tangherlini and Dray-'t Hooft shells we have seen above, as well as the ultra-boosted Kerr gyraton we turn to next.

\subsubsection{Ultra-boosted Kerr Gyraton}
The final example~\cite{Cristofoli:2020hnk} considers is the gyraton metric of a spinning particle. The gyraton metric 
\be
g_{\mu\nu}=\eta_{\mu\nu}-4G_N\alpha \log(|X^2-a^2|)q_\mu q_\nu\delta(q\cdot X)
\ee
describes ultra-boosted Kerr (or the metric surrounding a highly boosted spinning particle of mass $m$ and spin $s$) where 
$a^2$ gives the spin to mass ratio via
\be
a^\mu=\frac{1}{m}s^\mu,
\ee
$s^\mu$ is the spin vector, and $a\cdot q=0$.  The Kerr black hole can be obtained from the Schwarzschild solution via the Newman-Janis shift
\be
X^\mu\rightarrow X^\mu+i a^\mu,
\ee
which has reared its head in recent studies of the double copy~\cite{Arkani-Hamed:2019ymq,Cristofoli:2020hnk,Guevara:2020xjx}.  Here we see the ring singularity of Kerr at $X^2=a^2$, which extrudes a worldsheet (dashed curve in figure~\ref{support}) surrounding the worldline through the origin (bold).

We conclude this section with a couple of comments.  First, if we think of the gyraton as an ultra-boosted Kerr black hole as opposed to an approximate metric of an arbitrary spinning particle, we have a bound on $a\le 1$ in natural units. We see that any singular behavior is isolated to the hyperbolic slices within a Planck length of the lightcone. This worldsheet appears quite distinct from the geometry introduced in~\cite{Guevara:2020xjx}. The complexification of Minkowski space required in the Newman-Janus shift and employed in the worldsheet effective action of~\cite{Guevara:2020xjx}, however, lies at the heart of resolving collinear singularities of on-shell three point correlators~\cite{Pasterski:2017ylz}, and most likely presents a step towards establishing a worldsheet interpretation of the celestial sphere CFT. It would be interesting to pursue any connections that might manifest, now that we have identified this background as a generalized conformal primary state.

Second, the nature of this solution implies it should act as a canonical example of a source inserting spin memory~\cite{Pasterski:2015tva} at a point on the celestial sphere.  Its matter stress tensor should give a nonzero flux of angular momentum through null infinity.\footnote{Note the condition $a\cdot q=0$ does not get in the way of this fact.  In the rest frame of a spinning particle, the four momentum $(m,\vec{0})$ and spin vector $(0,\vec{s})$ are orthogonal. This remains true in the ultra-boost.  In the massless case, $s^\mu$ is replaced with the Pauli-Lubanski pseudovector and we measure the helicity of the particle moving with four momentum parallel to $q^\mu$.} Similar to the electromagnetic case~\cite{Pasterski:2015zua}, a single boosted particle will not radiate.  However, scattering between configurations that are approximately of this form at early and late times would be expected to give a non-zero radiative memory mode~\cite{Pasterski:2015tva,Himwich:2019qmj}, as opposed to just a change in the angular momentum aspect. 

\subsubsection{Superrotation Vacuum Transitions}
The Aichelburg-Sexl metric we began with is an example of a pp-wave.  Shock waves of the form studied by Dray and 't Hooft are relevant for inserting supertranslation hair.\footnote{See~\cite{Hawking:2016sgy} for the original proposal,~\cite{Donnay:2018ckb} for a near-horizon perspective and~\cite{Pasterski:2020xvn} for an examination of the effect of these modes on the experience of an infalling observer.  Note that we are looking at a planar shell here, as opposed to the spherical ones considered in those references.} Furthermore, we mentioned in the previous subsection that the gyraton should source spin memory, the canonical partner to the superrotation Goldstone. 
With these ideas in mind, it is natural to turn to another cut-and-paste proposal relevant to the asymptotic symmetry analysis.  In~\cite{Strominger:2016wns}, Strominger and Zhiboedov proposed interpreting superrotation vacuum-to-vacuum transitions as snapping cosmic strings.  Such geometries were further conjectured to generalize the celestial sphere to non-trivial topologies, which makes them of particular interest to celestial CFT with respect to possible constraints that could arise from modular invariance.

A metric which glues finite superrotations across the light-cone takes the form
\be\label{SZ}
\badat{2}
ds^2&=-du^2-2dudr+(2r^2\gamma_{z\bz}+\frac{1}{4}u^2\Theta(-u)(1+z\bz)^2\{\zeta,z\}\{{\bar\zeta},\bz\})dzd\bz\\
&\quad-ru\Theta(-u)(\{{\zeta},z\}dz^2+\{{\bar\zeta},\bz\}d\bz^2)\,,
\eadat
\ee
where
\be
\{\zeta,z\}=\frac{\zeta'''}{\zeta'}-\frac{3}{2}\Big(\frac{\zeta''}{\zeta'}\Big)^2.
\ee
We can massage this into the form of a classical double copy with generalized conformal primary metric form if we allow ourselves to complexify this metric.  It is straightforward to see that for the complexified case, by choosing
\be
\{\zeta,z\}=\frac{1}{(z-w)^4}\,,~~~\{{\bar \zeta},\bz\}=0\,,
\ee
which can be achieved with the finite complexified conformal transformation
\be
\zeta=\tan\frac{1}{\sqrt{2}(z-w)},~~~\bar{\zeta}=\bz\,,
\ee
the metric~\eqref{SZ} reduces to
\be
g_{\mu\nu}=\eta_{\mu\nu}+[\Theta(-X^2)\Theta(X^0)-1]\th^2_{\mu\nu}\,,
\ee
where $\th^2_{\mu\nu}$ is the superrotation Goldstone mode analyzed in~\cite{Donnay:2020guq} and $g_{\mu\nu}-\eta_{\mu\nu}$ is a generalized conformal primary metric of weight $\Delta=2$ and spin $J=2$ with Kerr-Schild vector $m_\mu$. We would also emphasize the difference between solutions of this form and solutions we would expect to source spin memory.  In Bondi gauge, the zero mode corresponding to spin memory should be paired with the superrotation Goldstone.  This observable is not, itself, a vacuum-to-vacuum transition between finitely superrotated solutions.

While concocting a complexified version to guarantee Kerr-Schild form is of interest from the point of view of this paper and looking at scattering around finite backgrounds (though one may need to go to (2,2) signature), sticking to infinitesimal transformations allows us to superimpose the complex conjugate mode, keep the metric real, and still find an application for our generalized conformal primary metrics, albeit not the Kerr-Schild double copy.  

So long as one sticks to linearized solutions, one can superimpose conformal primaries of various reference directions to build up more general angular profiles.  One could expect to straightforwardly construct the general form~\eqref{SZ}, which~\cite{Strominger:2016wns} showed matched Penrose's cosmic string metric, but the full C-metric bulk solution seems more subtle.  In~\cite{Luna:2018dpt}, it was shown that the C-metric can be written as a double copy. From the classification of source geometries in figure~\ref{support}, however, we do not expect the C-metric to be a single conformal primary.  It would be interesting to generalize our set up to other sourced and/or double copy examples. 

\section{Outlook}\label{sec:outlook}
We would like to conclude this paper by looking towards the future. Over the last few pages, we covered examples introduced decades ago by~\cite{Aichelburg:1970dh,Dray:1985ie,Ferrari:1988cc} in the interest of finding exact solutions to Einstein's equations.  Interestingly, recent literature like~\cite{Monteiro:2014cda,Arkani-Hamed:2019ymq,Cristofoli:2020hnk} is finally appreciating how these solutions arise from the double copy.  This change in method is powerful.  For our purposes, we are particularly interested in what it can teach us about a celestial CFT reorganization of scattering amplitudes.

In particular, let us point out that the method~\cite{Cristofoli:2020hnk} used makes the comparison here all the more interesting.  That paper showed how certain interesting exact solutions to the Einstein equations arose from an off-shell scalar-scalar-graviton 3-point function.  Here, we have shown that the metrics that appear in various interesting double copy examples are generalized conformal primary metrics. Combining this realization with the computation of~\cite{Cristofoli:2020hnk} seems to imply that we should also see this structure in the OPE coefficients of the celestial CFT.  It would be particularly interesting to pursue a celestial CFT interpretation of the method to reach these backgrounds rather than merely the metrics resulting therefrom.  

The goal of the Celestial Holography program is to gain insights into bulk physics and not just rewrite what is guaranteed from on-shell kinematics of scattering states. As such, we have sought to systematize our approach to building conformal primaries of different spin for radiative states, and generalize our formalism to non-radiative/off-shell states so that we gain more categories of candidate bulk physics interpretations for operators that might appear within celestial CFT OPE expansions and conformal block decompositions.

\section*{Acknowledgements}

We would like to thank Eduardo Casali, Laura Donnay, Alfredo Guevara, Cindy Keeler, Atul Sharma, and Herman Verlinde for discussions. The work of S.P. is supported by a fellowship at the Princeton Center for Theoretical Science.
The work of A.P. is supported by the European Research Council (ERC) under the European Union’s Horizon 2020 research and innovation programme (grant agreement No 852386).

\bibliographystyle{utphys}
\bibliography{shifting_spin}

\end{document}